\documentclass{aa}  
\usepackage{amsmath}
\usepackage{graphicx}
\usepackage{txfonts}
\usepackage{natbib} 
\usepackage{subfig}
\usepackage[T1]{fontenc}
\usepackage{subfig}
\usepackage[usenames, dvipsnames]{color}

\begin{document}

\def\mean#1{\left< #1 \right>}

   \title{The spectroscopic binaries RV\,Tauri and DF\,Cygni}
   \author{Rajeev Manick\inst{1,2}
          \and
          Devika Kamath\inst{3,4}
          \and
          Hans Van Winckel\inst{1}
          \and
          Alain Jorissen\inst{5}
          \and
          Sanjay Sekaran\inst{1}
          \and
          Dominic M. Bowman\inst{1}
          \and
          Glenn-Michael Oomen\inst{1,6}
          \and
          Jacques Kluska\inst{1}
          \and
          Dylan Bollen\inst{1,3}
          \and
          Christoffel Waelkens\inst{1}
          }

      \institute{Instituut voor Sterrenkunde (IvS), KU Leuven,
              Celestijnenlaan 200D, B-3001 Leuven, Belgium
              \and
              South African Astronomical Observatory, PO Box 9, Observatory, 7935, South Africa
              \and 
              Department of Physics and Astronomy, Macquarie University, Sydney, NSW 2109, Australia
              \and
	      Astronomy, Astrophysics and Astrophotonics Research Centre, Macquarie University, Sydney, NSW 2109, Australia
	      \and
	     Institut d'Astronomie et d'Astrophysique, Universit\'e Libre de Bruxelles, CP 226, Boulevard du Triomphe, 1050 Bruxelles, Belgium
	     \and 
	     Department of Astrophysics/IMAPP, Radboud University, P.O. Box 9010, 6500 GL Nijmegen, The Netherlands\\
              \email{rajeevmnck@gmail.com}
             }

   \authorrunning{Manick et al. 2018}
   \titlerunning{The spectroscopic binaries RV\,Tau and DF\,Cyg}

   \abstract
   {Some RV Tauri stars show a long-term photometric variability in their mean magnitudes. DF\,Cygni (DF\,Cyg), the only RV Tauri star in the original \textit{Kepler} field,
   and the prototype RV\,Tauri (RV\,Tau) are two such stars. 
   }
   {The focus of this paper is on two famous but still poorly understood RV Tauri stars: RV Tau and DF Cyg. 
   We aim at confirming their suspected binary nature and deriving their orbital elements to investigate the impact of their orbits 
   on the evolution of these systems. This research is embedded into a wider endeavour to study binary evolution of low- and intermediate-mass stars. 
   }
   {The high amplitude pulsations were cleaned from the radial-velocity data to better constrain the orbital motion, allowing us to obtain accurate orbital parameters. 
   We also analysed the photometric time series of both stars using a Lomb-Scargle periodogram.
   We used Gaia DR2 parallaxes in combination with the SEDs to compute their luminosities. These luminosities were complemented with the
   ones we computed using a period-luminosity-colour (PLC) relation for RV\,Tauri stars.
   The ratio of the circumstellar infrared (IR) flux to the photospheric flux obtained from the SEDs was used to estimate the orbital inclination of each system. 
   
   }
   {DF\,Cyg and RV\,Tau are binaries with spectroscopic orbital periods of 784\,$\pm$\,16 days and 1198\,$\pm$\,17 days, respectively. These orbital periods
   are found to be similar to the long-term periodic variability in the photometry, indicating that binarity indeed explains the long-term photometric variability.  The SEDs of these systems indicate the presence of a circumbinary disc.
   Our line-of-sight grazes the dusty disc, which causes the photometric flux from the star to extinct periodically with the orbital period.
   Our derived orbital inclinations enabled us to obtain accurate companion masses for DF\,Cyg and RV\,Tau and these were found to be 0.6\,$\pm$0.1 M$_\sun$ and 0.7\,$\pm$\,0.1 M$_\sun$, respectively.
   The derived luminosities suggest that RV\,Tau is a post-AGB binary, while DF\,Cyg is likely a post-RGB binary. 
   Analysis of the \textit{Kepler} photometry of DF\,Cyg revealed a power spectrum with side lobes around the fundamental pulsation frequency. This modulation corresponds to the spectroscopic orbital period and hence to the long-term photometric period.
   Finally we report on the evidence of high velocity absorption features related to the H$_\alpha$ profile in both objects, indicating outflows launched 
   from around the companion.
   }{}


   \maketitle
%

\section{Introduction}
RV Tauri stars are large-amplitude pulsators that occupy the high-luminosity end of the population II Cepheids \citep{wallerstein02}.
They are primarily defined by their lightcurve characteristics which display alternating deep and shallow minima. 
The period between two consecutive deep minima is also termed as the `formal period' while the `fundamental period' is defined as half of the former.
The formal period has commonly been quoted as ranging from 30 to 150 days \citep[e.g.,][]{fokin94,pollard97,wallerstein02,percy07}. However, 
the lightcurve analyses of a sample of population II Cepheids in the SMC and LMC by \citet{soszynski08,soszynski10} showed that the shortest 
pulsation period among RV Tauri stars is around 20 days.

Apart from the pulsational variability, a number of RV Tauri stars display a long-term periodic variability in their mean flux of order 600 to 2600 days.
These stars have been classified photometrically as RVb types, while the ones with a constant mean flux in their 
time series are classified as RVa photometric types \citep{kukarkin69,evans85,pollard96}.

Since the discovery of the long-term variability in the photometric time series of several RV Tauri stars, there have been uncertainties about the mechanism giving rise to it. 
Attempts have been made to explain the RVb phenomenon using two hypotheses: 1) a mechanism linked to the intrinsic variability of the star \citep[e.g.,][]{tsesevich75,zsoldos96}
and 2) the binary hypothesis whereby the primary is being obscured periodically by dust \citep[e.g.,][]{percy93,fokin94,pollard97,vanwinckel99}.

Given that there is a strong correlation between the presence of circumstellar dust and the appearance of the RVb phenomenon among RV Tauri stars \citep[e.g.,][]{lloydevans85}, 
it has become clear during the last few decades that the appearance of the RVb phenomenon is related to the dust and binary nature of the central star
\citep{waelkens91a,waelkens95,vanwinckel99,manick2017}. The RVb class is best explained by a configuration in which there is periodic obscuration of the orbiting primary star by a dusty disc that surrounds the binary system. 
The dust disc is inclined with respect to the observer in such a configuration that at inferior conjunction the obscuration of the primary is at its maximum, while at superior conjunction the obscuration of the primary is minimum. 

The two stars included in this study are RV\,Tauri (RV\,Tau) and DF\,Cygni (DF\,Cyg). These stars belong to the RVb class and have a 
broad IR excess in their SEDs \citep{deruyter06}. It is now well established that this type of SED points to the presence of a stable circumbinary disc.
For a review of all the observational evidence, we refer to \citet{vanwinckel2018}.

RV\,Tau itself is the prototype based on which all RV Tauri variables have been classified \citep{ceraski1905}. The lightcurve of RV\,Tau shows pulsations with a formal period 
of around 78.5 days and a fundamental period half of that \citep{preston63,dawson1982}. Photometric analyses of RV\,Tau by \citet{percy97} showed that this 
variation is not purely periodic. The depths of minima display irregularities from cycle to cycle. They suggest that the irregularity is caused either 
by an intrinsic instability in the pulsation or is due to the presence of a secondary period beating against the primary period on short timescales. 
The long-term lightcurve of RV\,Tau displays a long periodic variation of around 1200 days in the mean brightness \citep{lloydevans85}. 

DF\,Cyg is particularly interesting because it is the only RV\,Tauri star
in the original {\it Kepler} field \citep{borucki2010,bodi2016}. The outstanding quality of {\it Kepler} data available for this star has made it subject to a number of recent studies.
\citet{bodi2016} have compiled 50 years of AAVSO data and around 4 years of {\it Kepler} data to produce a detailed analysis of the lightcurve of DF\,Cyg. 
The formal pulsation period is 49.85 days, the fundamental period being half of this (i.e, 24.9 days). The long period is of the order of 779 days in the {\it Kepler} data.
The main results \citet{bodi2016} derive from their study is the detection of a rich set of subharmonics in the Fourier series and also the presence of strong non-linear 
effects in the pulsations revealed by the {\it Kepler} time series.

\citet{vega2017} analysed the raw fluxes of DF\,Cyg to find that the overall mean flux decreases by around 90 percent between long-period maximum and long-period minimum.
Interestingly, the percentage decrease in the mean flux is exactly the same in the pulsation amplitude from the high-extinction to low-extinction phase.
They conclude that this can be fully explained using a scenario in which the intrinsic pulsational amplitude of the star is constant and the
star is being obscured by dust through the RVb cycle. These results are complementary to those of \citet{kiss2017}, who showed that the mean brightness of the star
correlates with the pulsational amplitude, which is again best explained by now widely known as the ``dust obscuration model''. 
RV\,Tau and DF\,Cyg have been postulated to be binaries either because of their RVb nature \citep[e.g.,][]{bodi2016,vega2017} or because of the presence of a disc \citep{deruyter05}. 

The paper is ordered as follows. In Section \ref{section:data1} we describe our photometric and spectroscopic data for the two stars. 
In Section \ref{section:photometricanalysis}, we outline the time series photometric analysis of the two stars.
The analysis of the spectroscopic data to obtain the stellar parameters and the radial velocities is described in Section \ref{section:specanalysis}. 
In Section \ref{section:luminositydf}, we describe the methods used to constrain the luminosities.
We describe the orbital 
analysis for each star in Section \ref{section:orbitalanalysis}. The results of the paper are discussed in Section \ref{section:discussiondf}. 
We conclude in Section \ref{section:conclusiondf}.

\section{Data}\label{section:data1}
\subsection{Photometric time series data} \label{section:phot}
\subsubsection{RV\,Tau}
We gathered $V$-band photometric time series for RV\,Tau from the All Sky Automated Survey \citep[ASAS,][]{pojmanski02} database (ACVS\footnote{http://www.astrouw.edu.pl/asas/?page=acvs}).
ASAS is a photometric monitoring campaign that mapped the sky south of $\delta$ = $+$28$^{\circ}$ to look for variable stars brighter than a $V$-mag of 14. 
The observations are carried out at two sites, Las Campanas Observatory (Chile, since 1997) and Haleakala (Maui, since 2006).
Each telescope is fully automated and equipped with two wide-field instruments, which can observe simultaneously in the $V$- and $I$-bands. 

The ASAS pipeline data-reduction 
routine performs the aperture photometry into five different aperture size measurements ranging 
from 2 pixels to 6 pixels (MAG$_0$, MAG$_1$, MAG$_2$, MAG$_3$ and MAG$_4$). The photometry of the target star is not contaminated if there is another star more than 15 arcsec away.
We checked the field of RV\,Tau and the closest star is IRAS 04442$+$2558 at a separation of $\sim$ 463 arcsec.
Thus, with $V$=9.8 for RV\,Tau, we used data from the largest aperture size of 6 pixels (MAG$_4$), which includes most light from the star and is free from blending effects. We limited our photometric analysis to the A-grade data quality.
The final photometric dataset from ASAS consists of 180 flux points with a timespan of $\sim$ 2537 days.

\subsubsection{DF\,Cyg}
To construct optimised {\it Kepler} light curves for DF\,Cyg (KIC~7466053), we downloaded the target pixel files from the Mikulski Archive for Space 
Telescopes (MAST\footnote{MAST website: http://archive.stsci.edu/kepler/}) and we created custom masks for each long-cadence data quarter following 
the methodology described in \citet{papics2017}. This minimized long-period instrumental trends since we included more pixels in the mask compared 
to the default MAST masks, being careful to avoid nearby possible sources of contamination. Then we merged all the data quarters into a single light curve by removing  
obvious outliers. The final dataset consists of 65265 datapoints, with each flux point taken every $\sim$ 30 min, for around 1471 days.

\subsection{SED data}
To construct the SEDs, we gathered different sets of broad-band photometry available in the literature.
We excluded any photometric point that was marked in the catalogue as having an unreliable flux determination. 
In the two sections below, we give a description of the SED data for the two stars.

\subsubsection{RV\,Tau}
The photospheric part of the SED is characterised by $UBVRI$ photometry from the Johnson \citep[$UBVR$,][]{johnson1953}, and Sloan Digital Sky Survey 
(SDSS) $I$-band filters \citep{york2000}. These were complemented with additional 
$B$- and $V$-band photometry from the TYCHO catalogue \citep{hog2000}.
The near-IR region is covered by $J$-, $H$-, and $K$-band 
photometry from the Two Micron All Sky Survey (2\,MASS, \citealt{skrutskie06}) and Johnson $J$- and $H$-band photometry from the Catalogue of Stellar Photometry \citep{ducati2002}. 
The mid-IR is composed of fluxes from the WISE and AKARI catalogues \citep{murakami2007,wright10,ishihara2010}.

\subsubsection{DF\,Cyg} \label{section:sed_data_dfcyg}
Most of the SED photometric fluxes of DF\,Cyg have been obtained from the same catalogues as those of RV\,Tau, with the exception of 
GALEX UV fluxes from MAST$^{1}$ and the far-IR region complemented with data from IRAS \citep{helou88}.

We noticed that the 2MASS flux points did not follow the general SED trend because they were taken at a phase when the star was highly extinct. 
These fluxes were corrected to their unextinct values by \citet{vega2017}, using the extinction models of \citet{cardelli89}. We used these corrected fluxes in our SED analysis of DF\,Cyg.

\subsection{Spectroscopic data} \label{section:spec}
RV\,Tau and DF\,Cyg have been monitored spectroscopically using the semi-automatic 1.2\,m Flemish Mercator telescope,
installed at the Roque de los Muchachos Observatory in La Palma, Spain.
The telescope is equiped with a high-resolution fibre-fed echelle spectrograph (HERMES).
The high-resolution mode is operated using a high-resolution fibre (HRF) that offers a spectral resolution of R $\sim$ 85\,000 at a 2.5'' sky aperture. 
In HRF mode and under good observing conditions (1'' seeing), HERMES reaches a S/N of 100 at 5500\,$\AA$ for a $V$=10.4 source in a 1-hour exposure.
At the end of each observing night, the data are reduced through a dedicated pipeline \citep[see][for details]{raskin11}.

We obtained a total of 140 spectra for RV\,Tau and 83 spectra for DF\,Cyg, with a total timespan of $\sim$ 3100 and 3000 days, respectively.
The spectra were taken between July 2009 and March 2018.

\section{Photometric analysis} \label{section:photometricanalysis}

\subsection{Timeseries analysis of RV Tau}
Prior to analysing the time series, we excluded any extreme outliers in the data, which were likely due to instrumental effects. This was done by excluding any 
flux value beyond $\sim$ 4\,$\sigma$ from the mean. The dominant pulsation frequencies were then obtained from a Lomb-Scargle periodogram of the time series \citep{scargle82}. 
We treated a frequency peak as being significant if it had an S/N of $\sim$\,4 or higher in the Lomb-Scargle periodogram \citep{breger93,kuschnig97}. Table \ref{table:RV_Tau_ASAS} displays 
all the significant periods we found and their respective amplitudes and phases. Column 4 in the same Table lists the period assignments. The long-term variability 
is denoted as the orbital period $P_{\rm orb}$ while the others are harmonics of the fundamental period ($P_0$). We find 
two other periods namely, 45.7 days and 91 days which are significant in the periodogram, but are not multiples of $P_0$.

\begin{table}
  \centering
  \tiny
  \begin{tabular}{@{} lcccc @{}}
    \hline \hline
    period    &  amplitude   &  characteristic     &  phase       &     S/N \\   
    (days)    &  (mag.)      &               &        (rad)      &         \\
    \hline             
 39.3{$^*$}   &  0.45        &   P$_0$       & -0.09             &     8.0 \\ 
 1290.0{$^*$} &   0.21       &   P$_{\rm orb}$            &  0.15        &     5.9 \\ 
 91.0{$^*$}   &  0.14        &    -           & -0.16        &     4.3 \\ 
 79.1{$^*$}   &  0.12        &  2\,P$_0$           &  0.30        &     4.4 \\ 
 98.6         &  0.10        &  5/2\,P$_0$             & -0.25        &     4.0 \\ 
 19.6{$^*$}   &  0.08        &  P$_0$/2             & -0.48        &     4.1 \\ 
 45.7         &  0.08        &    -           & -0.23        &     4.0 \\ 
      \hline
   \end{tabular}
  \caption{Significant periodicities found in the ASAS photometry of RV\,Tau. The periods marked with an asterisk ${^*}$ were also found in the radial velocities.}
  \label{table:RV_Tau_ASAS}
\end{table} 

\begin{figure}
   \centering
   \includegraphics[width=9cm,height=8cm]{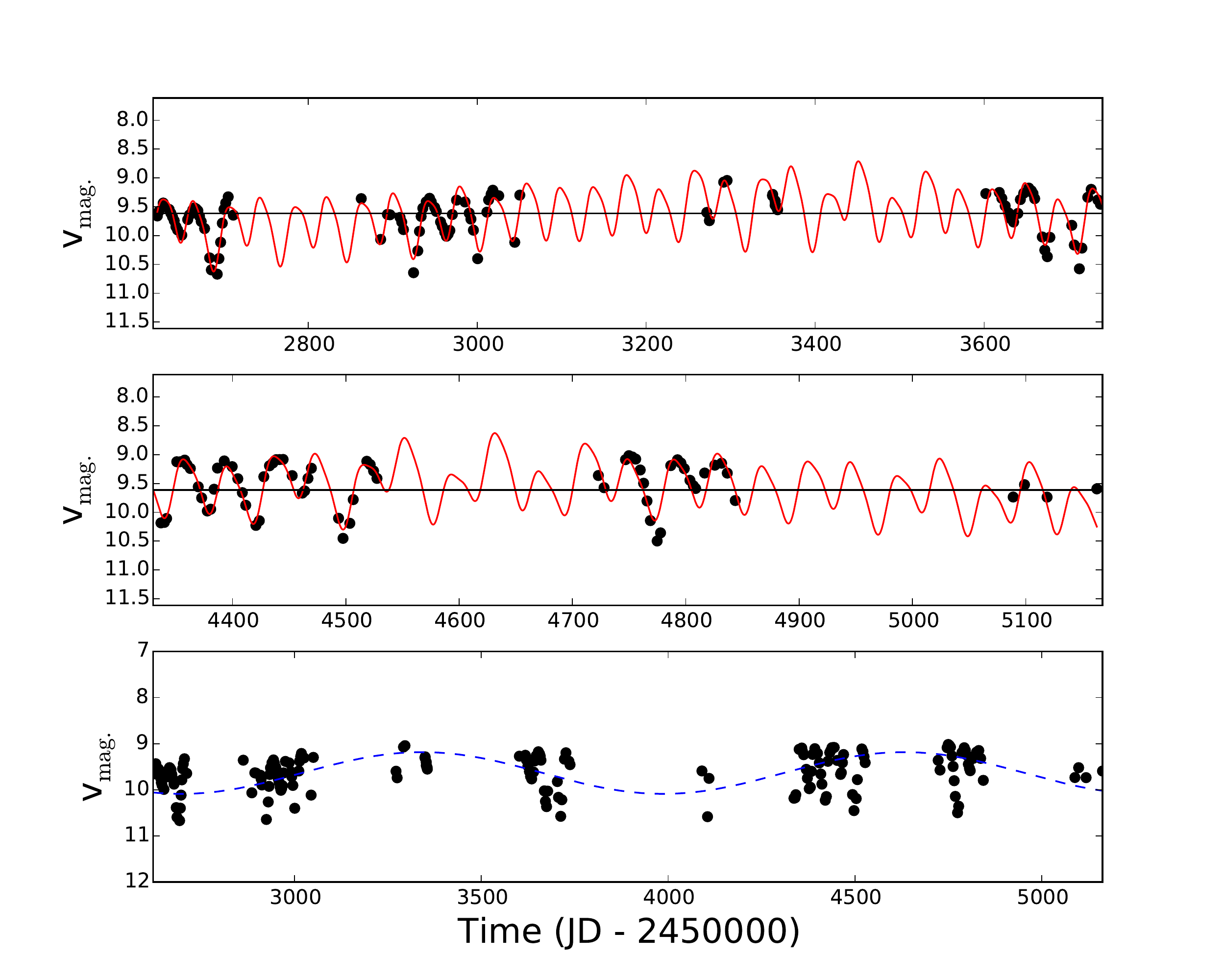}
   \caption{The upper two panels show the pulsation variability and the bottom panel displays the RVb variation in the ASAS lightcurve of RV\,Tau.}
     \label{figure:rvtau_rvb}
\end{figure} 

\subsection{Timeseries analysis of DF Cyg}   
The {\it Kepler} time series of DF\,Cyg has been analysed in details by \citet{bodi2016}, \citet{vega2017} and \citet{kiss2017}.
The results of \citet{bodi2016} show a number of sub-harmonics in the {\it Kepler} data, while \citet{vega2017} and \citet{kiss2017} conclude 
that there is a substantial modulation of the pulsation amplitude during the RVb cycle.

\subsubsection{Modulation in Fourier series}
\begin{figure}
\centering
 \includegraphics[width=9cm,height=7cm]{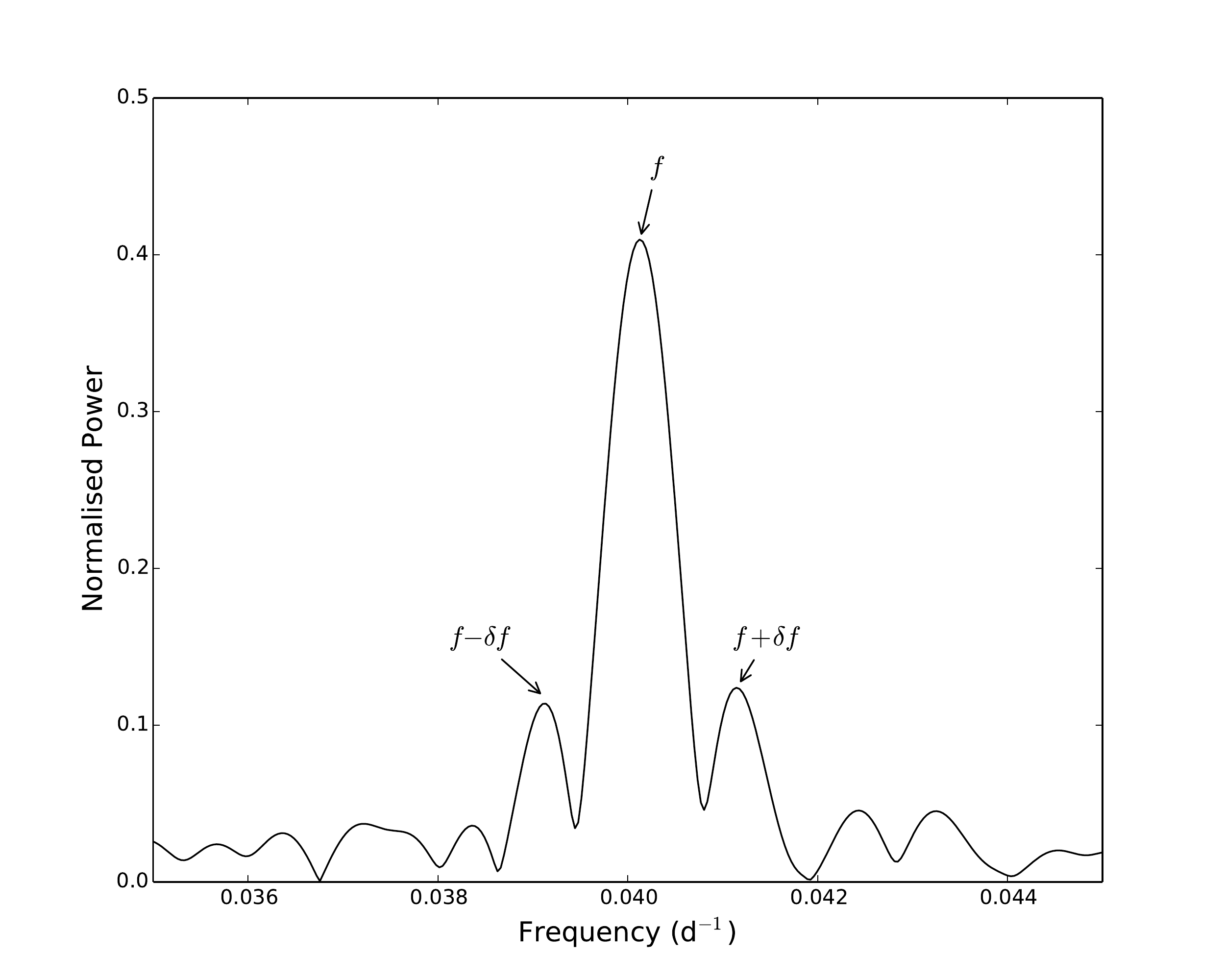}
\caption{Fourier spectrum of DF\,Cyg showing the fundamental peak frequency ($f$) corresponding to a fundamental period of 24.9 days.
The sidelobes at $f + \delta f$ and $f - \delta f$  indicate the pulsational frequency modulated due to orbital motion.}
\label{figure:fourier_orb_removed}
\end{figure}

Upon closer inspection of the Fourier periodogram of DF Cyg obtained using four years of \textit{Kepler} data, we found that the fundamental frequency appeared to be modulated with peaks at 
slightly higher and lower frequencies than the intrinsic pulsation period of DF\,Cyg (see Fig. \ref{figure:fourier_orb_removed}). The frequency separation is found to be $\sim$ 0.00128 c/d, which indicates periodic modulation with a period of P\,=\,778$\pm$8 day. We note that this periodic modulation is within errors of the binary orbital period derived from spectroscopy (see Sect. \ref{section:orbitalanalysis}) which also corresponds to the longterm RVb period of 779 days \citep{bodi2016}. Hence the modulation of the fundamental period could be caused by either periodic obscuration of the star by the circumbinary dusty disc, or the orbital motion of the binary system.

\subsubsection{Extinction through the disc} \label{section:extinctionvariation}
\begin{figure}
   \centering
   \includegraphics[width=9cm,height=7cm]{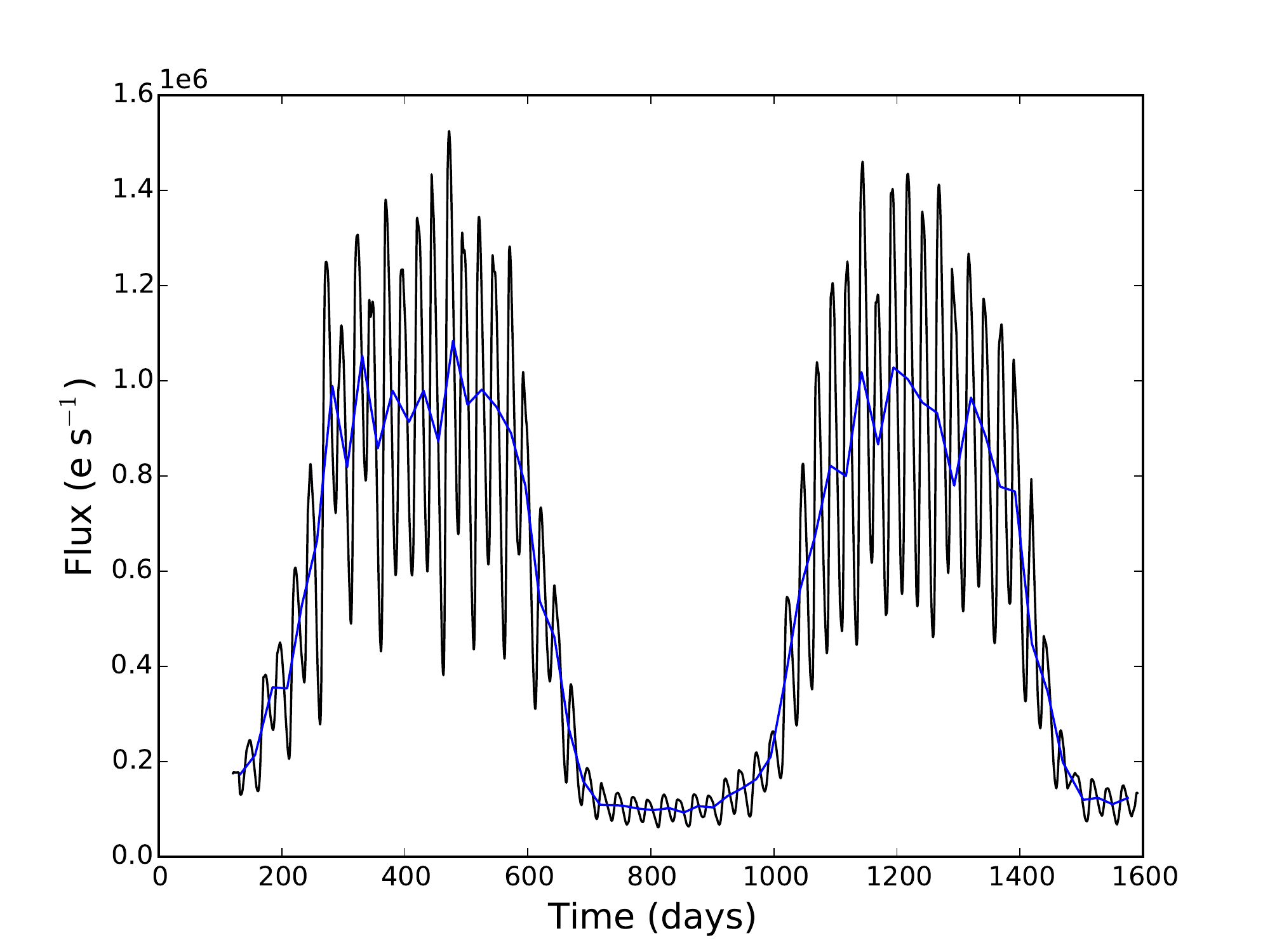}
   \includegraphics[width=9cm,height=7cm]{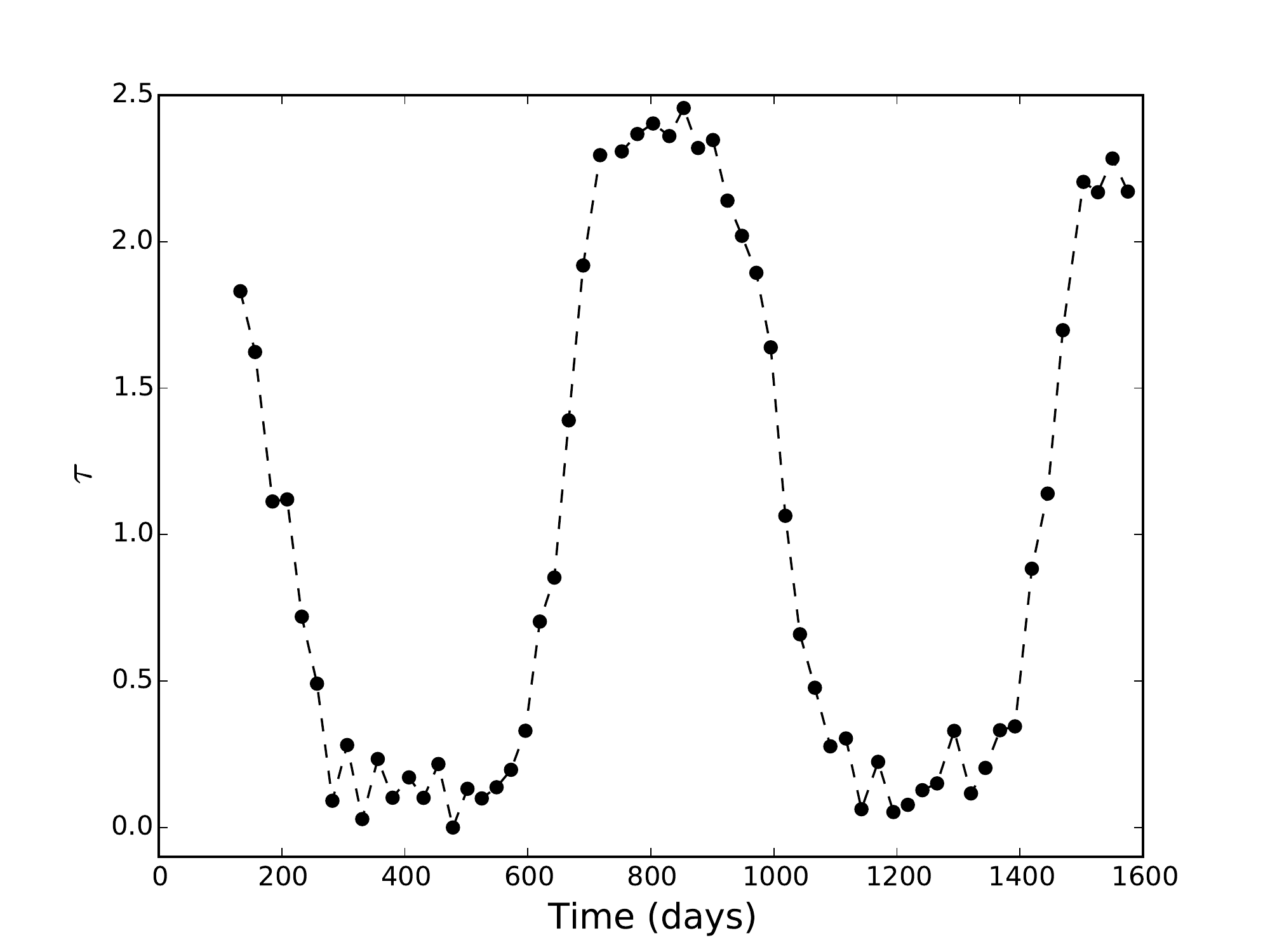} 
   \caption{Top: The mean flux variation throughout the RVb cycle (blue curve). Bottom: the variation of the opacity $\tau$ through the RVb cycle.}
     \label{figure:extinction_variation}
\end{figure}

The RVb variability indicates that the flux from the star is periodically extinct (see top panel of Fig. \ref{figure:extinction_variation}). 
The dust grains in the line-of-sight attenuate the intensity of the incoming photons by an amount which can be expressed in flux through the following equation \citep[e.g,][]{carroll96}:

\begin{equation} \label{equation:extinction}
I_{\lambda} = I_{\lambda, 0}\,{\rm exp}(-\tau_{\lambda})\,,
\end{equation}

\noindent where $I_{\lambda, 0}$ is the unattenuated flux and $\tau$ is the optical depth which depends on the physical properties and structure of the disc at wavelength $\lambda$.
We used the {\it Kepler} raw fluxes and Eq. \ref{equation:extinction} to investigate the variation of the optical depth through the RVb cycle.

We divided the time series into 59 chunks, with each chunk corresponding to one full pulsation cycle and calculated 
the mean of each cycle, which is shown in the upper panel of Fig. \ref{figure:extinction_variation} by the blue line.
Assuming that the maximum flux from the star is the unattenuated flux ($I_0$), we calculated the attenuated flux 
in the $V$-band ($I_V$), which was used to obtain the variation of the optical depth as a function of time. 
The result is displayed in the lower panel of Fig. \ref{figure:extinction_variation}.

A change of $\tau$ from 0 up to a value of $\sim$ 2.5 indicates that the material in our line-of-sight is optically thick at the phase when the star is fully extinct compared to the phase when the extinction is minimal.
In addition, we note from the lower panel of Fig. \ref{figure:extinction_variation} that the variation of $\tau$ through the orbit is relatively smooth. 
This implies that the material being grazed by our line-of-sight does not contain local regions of high attenuation (like clumps or disc inhomogeneities).

\section{Spectroscopic analysis} \label{section:specanalysis}

\subsection{Radial velocities} \label{section:rvs}
The individual radial velocities are obtained by cross-correlating the extracted spectrum with a mask of individual lines adapted to the stellar spectral type. 
For both RV\,Tau and DF\,Cyg we used a G2 mask, which has around 2200 individual lines identified. With a good overall S/N in the spectrum, the cross-correlation
profile may be considered as a mean line profile.
Each radial-velocity is computed by taking the mean and standard-deviation of a Gaussian fitted to this mean profile. 
The zero point velocity is calibrated using IAU radial-velocity standards \citep[see][for details]{raskin11}.

\subsection{Atmospheric differential motion and shocks} \label{section:shocks}
\begin{figure}
   \centering
   \includegraphics[width=8cm,height=6cm]{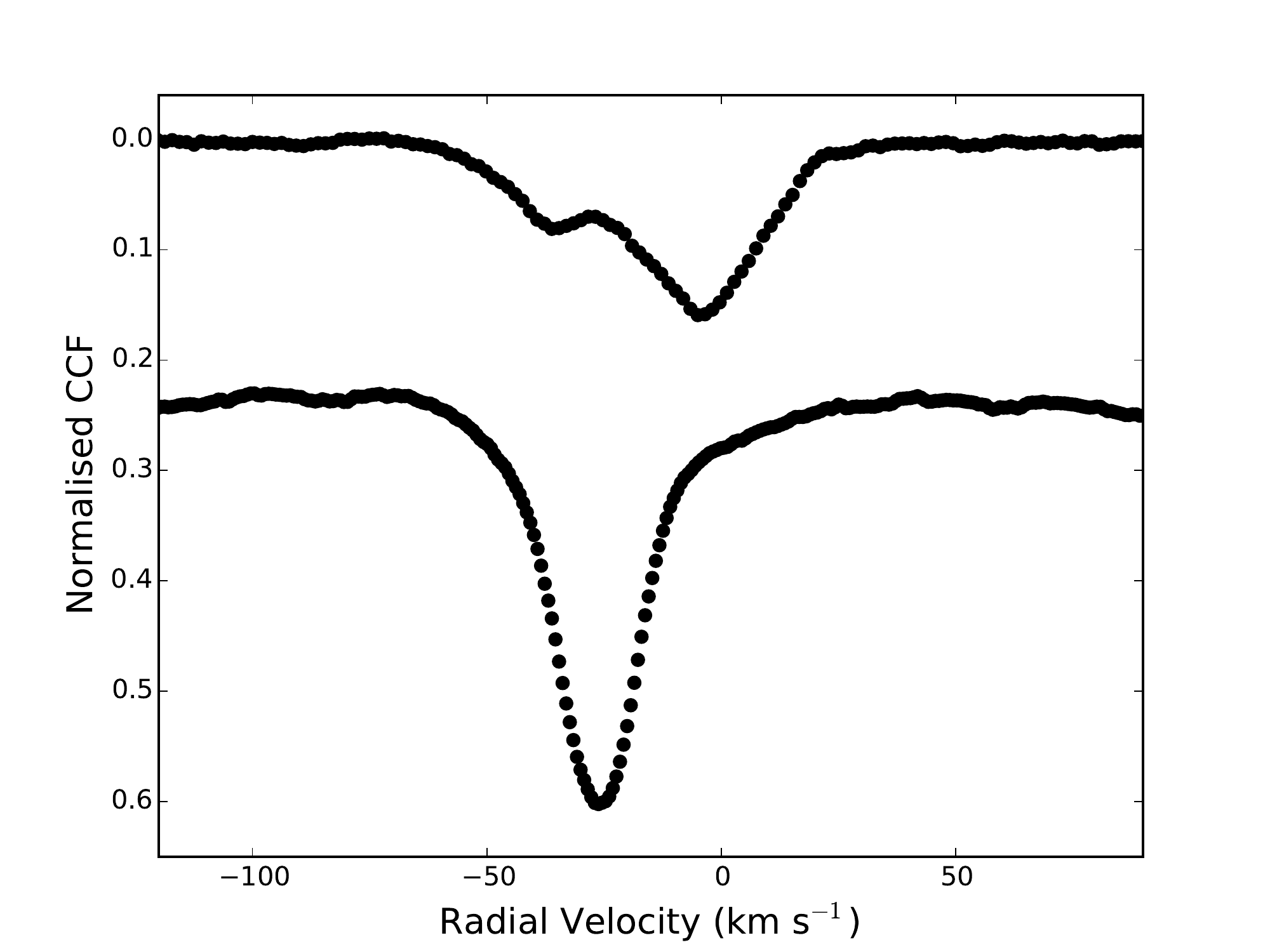}
   \caption{Top: a double-peaked CCF profile displaying a blue- and red-shifted component, which is a signature of shock. Bottom: a CCF profile which is unaffected by shocks.
}
     \label{figure:ccf}
\end{figure}

The shape of the cross correlation function (CCF) profile is representative of the overall line profile of individual lines in a given spectrum. At phases when the stellar atmosphere is stable 
(i.e., no shocks), the CCF is a Voigt profile centered at a radial-velocity which traces the orbital motion and pulsation only (see bottom plot of Fig. \ref{figure:ccf}). 
On the other hand, the presence of differential atmospheric motion in the photosphere of pulsating stars results into an overall mean CCF profile which is distorted or double peaked (see top plot of Fig. \ref{figure:ccf}). 

\begin{figure}
   \centering
   \includegraphics[width=8cm,height=6cm]{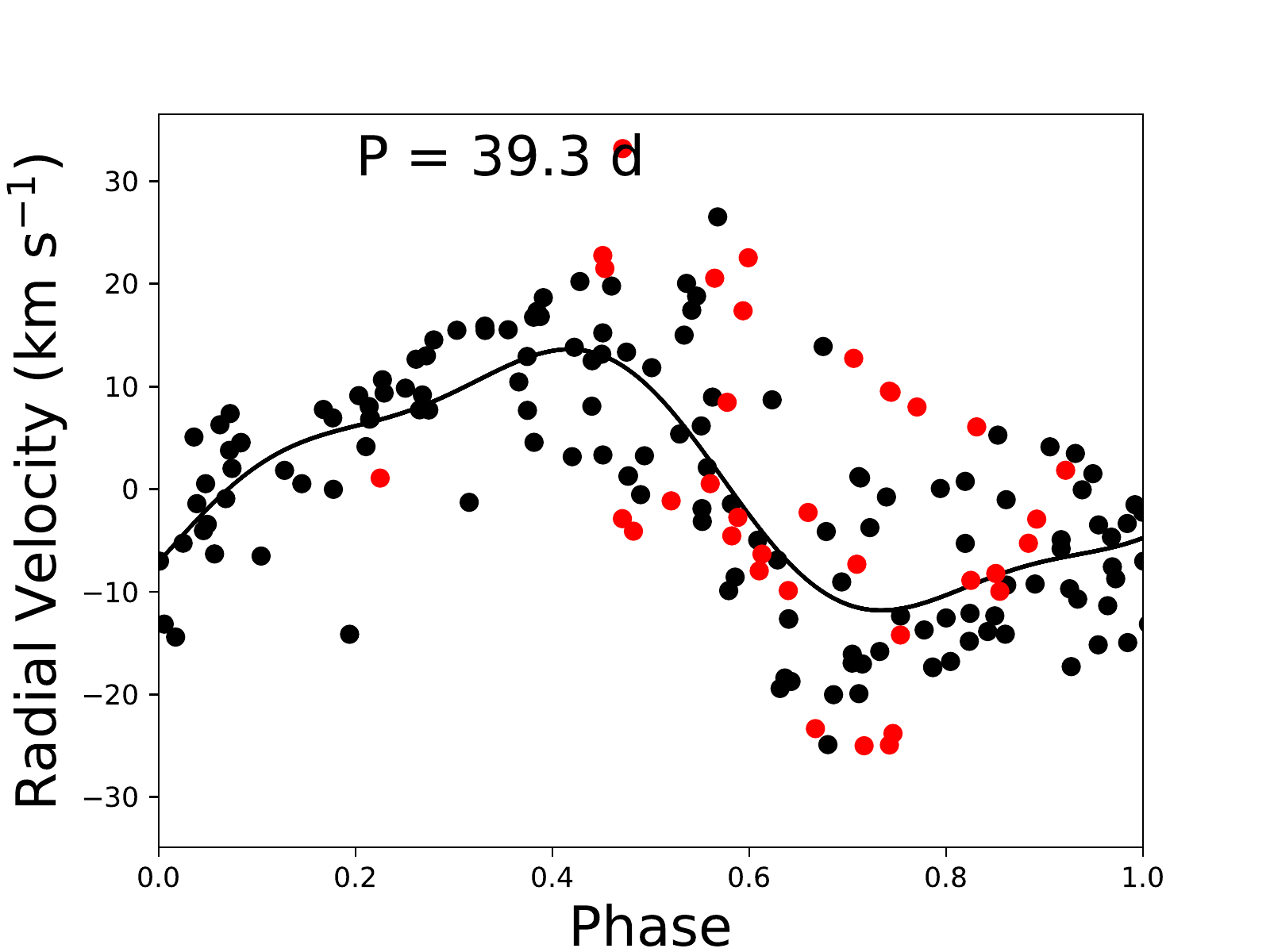}
   \includegraphics[width=8cm,height=6cm]{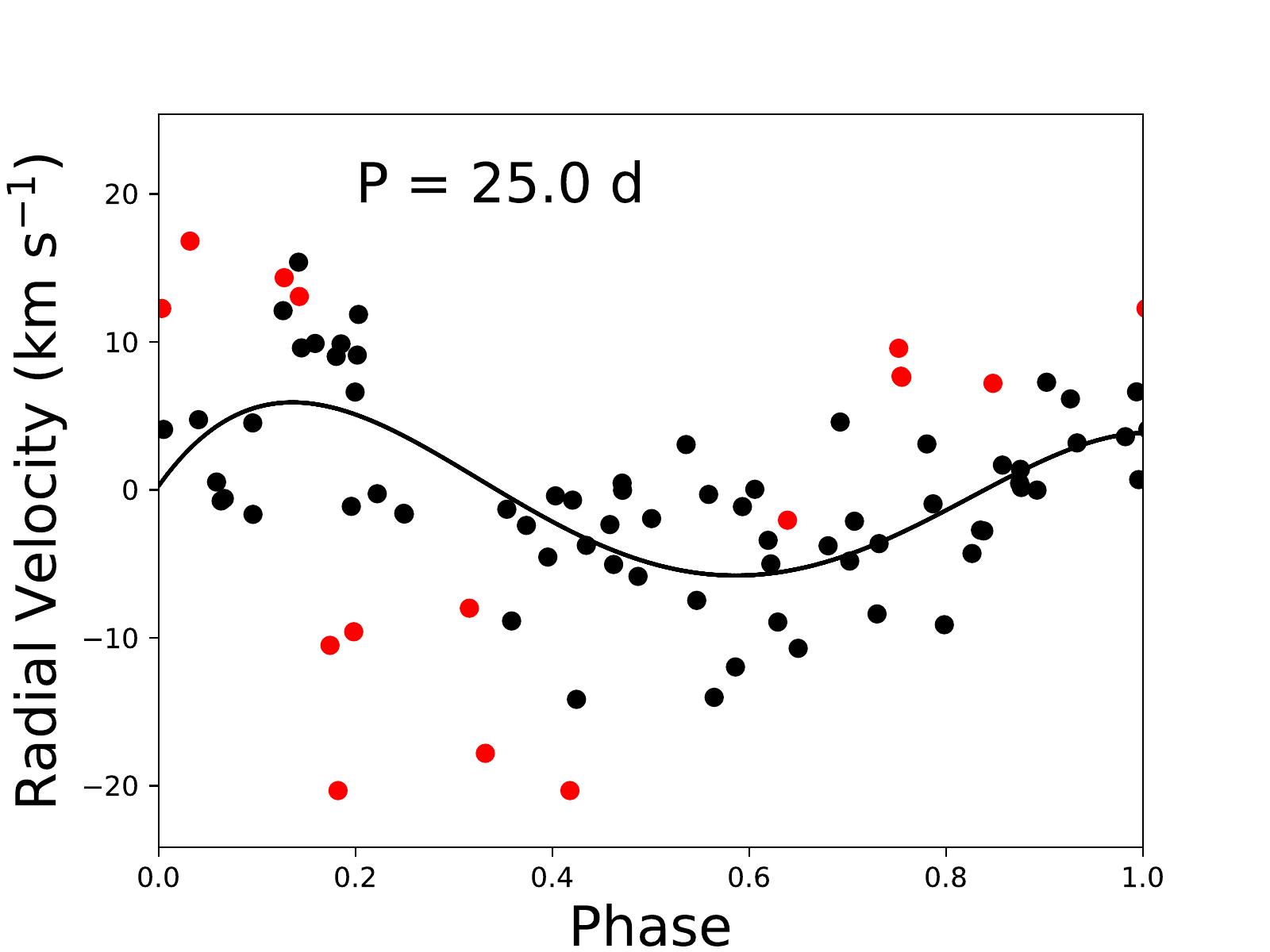} 
   \caption{Radial-velocities phased by the pulsation period. The red dots are RVs that are related to shocks. Top panel: RV\,Tau, Bottom panel: DF\,Cyg. 
   The black line represents the sinusoidal model (fundamental mode and harmonics) fitted to the data.}
     \label{figure:shock_puls_phase}
\end{figure}

In our orbital analysis, we excluded 35 spectra for RV\,Tau and 15 spectra for DF\,Cyg, which are related to the presence of these line distortions in the spectra.
The radial velocities corresponding to these spectra are plotted as red dots in Fig. \ref{figure:shock_puls_phase} to show the pulsation phase at 
which they occur. These radial velocites were not included in the orbital analysis.

Differential motions are known to occur at specific phases of the pulsation cycle \citep{baird84}. 
While this is indeed the case for RV\,Tau, in which it happens mostly at a phase when the star is expanding,
we note that the distorted line profiles appear at almost all phases for DF\,Cyg (see bottom panel of Fig. \ref{figure:shock_puls_phase}).

\begin{figure}[h!]
   \centering
   \includegraphics[width=8cm,height=6cm]{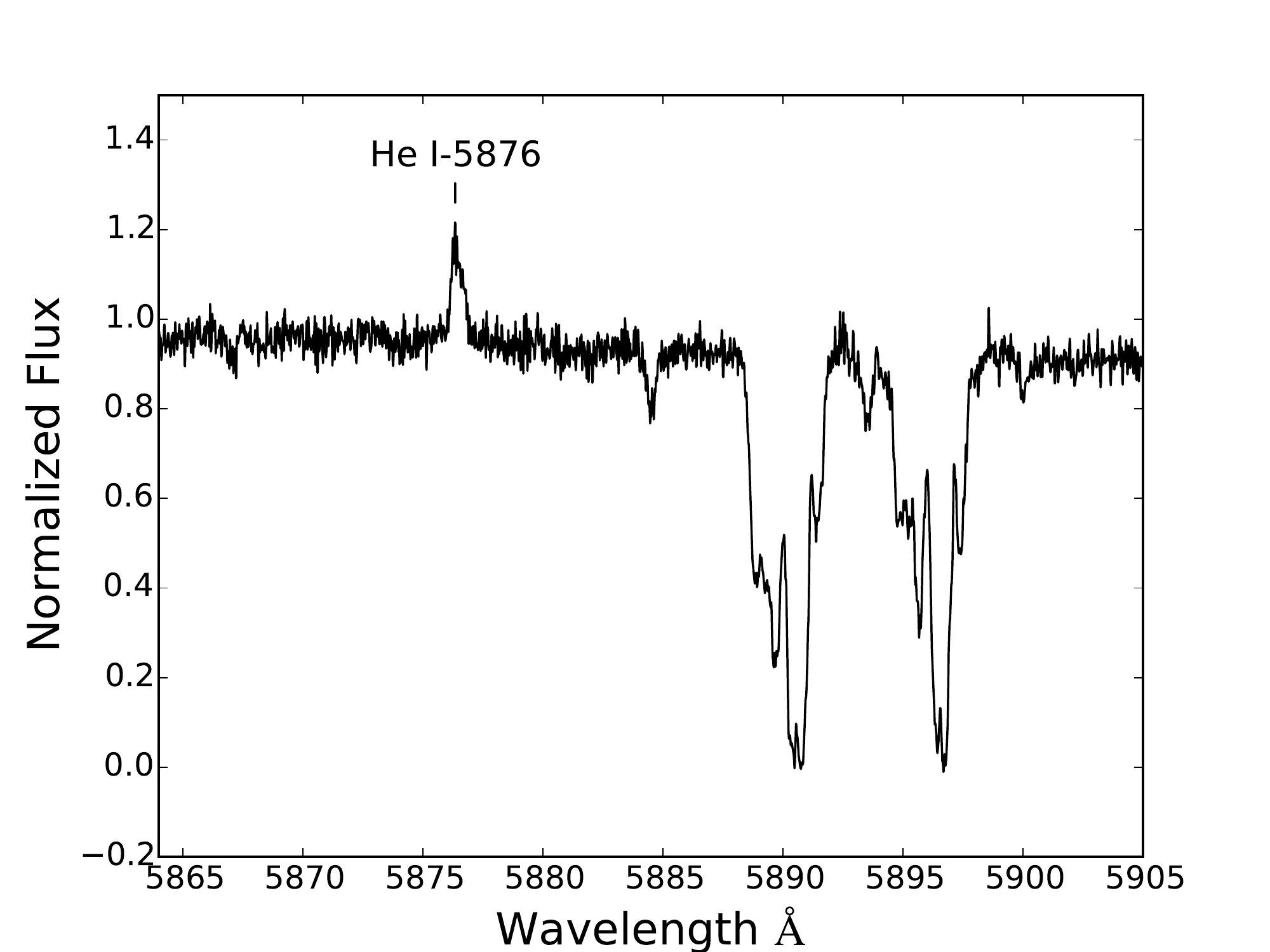}
   \includegraphics[width=9.5cm,height=7.5cm]{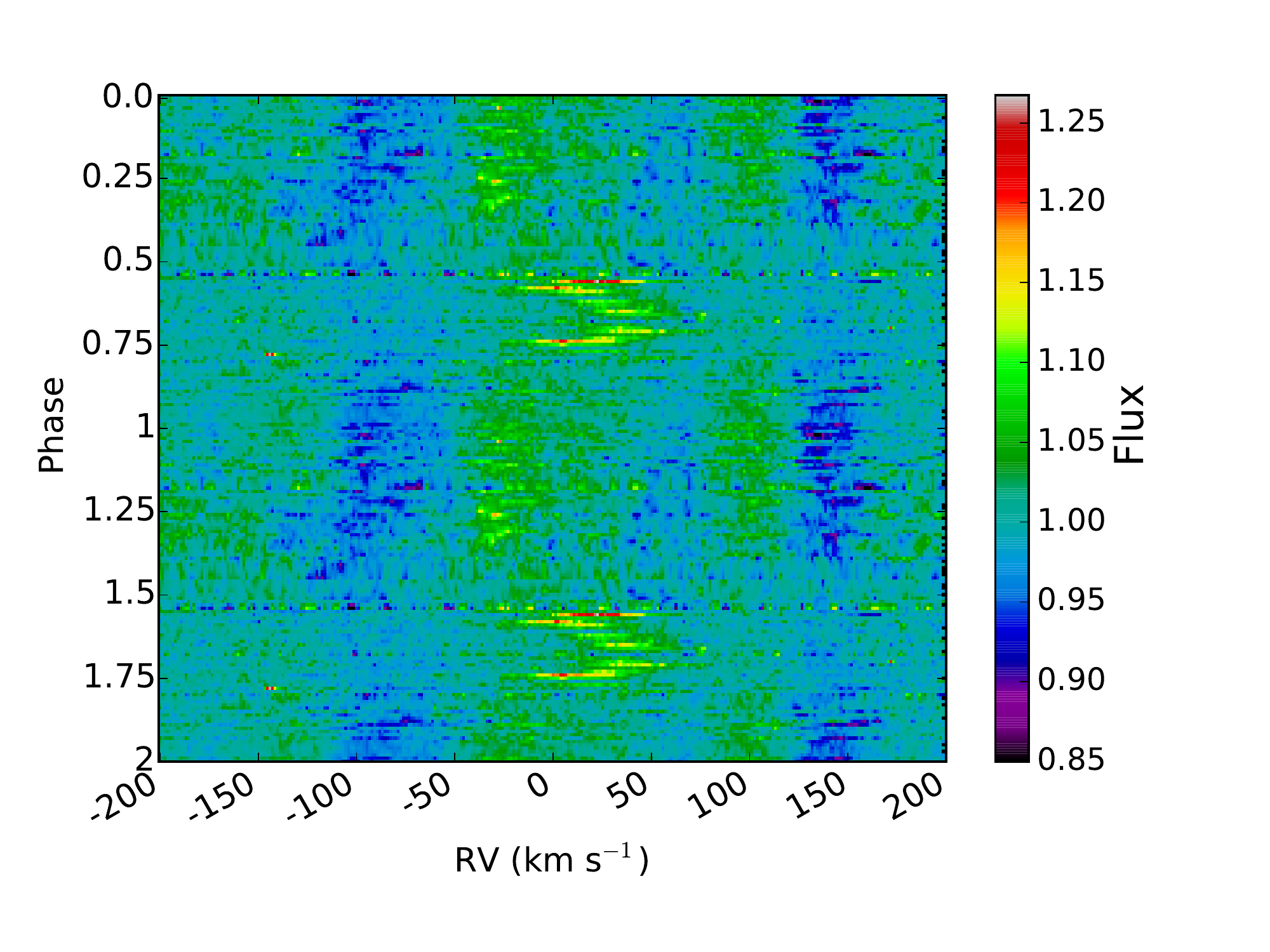}
   \caption{Upper panel: An example of the appearance of the He I-5876 line at a phase when a double-peaked CCF occurs in RV\,Tau. Lower panel: A dynamical plot of all the spectra of RV\,Tau 
   around the HeI-5876 line folded on the fundamental pulsation period of 39.3 days. }
     \label{figure:spectra_shock}
\end{figure}

These differential motions occur mostly in the upper stellar atmosphere and are known to result in shocks when there is a violent collision between oppositely moving layers \citep{baird84}.
One of the tracers of these shocks is H$\alpha$ emission \citep{gillet90,pollard97}. However, a more direct evidence of strong shocks is
the presence of the He\,I-5876 line in emission \citep{maas05}. We searched for the appearance of this line in the spectra of RV\,Tau and DF\,Cyg.
While RV\,Tau displays the He\,I-5876 line in emission at some phases (see top panel of Fig. \ref{figure:spectra_shock}), this line is not seen in DF\,Cyg. 

We have also included in the lower panel of Fig. \ref{figure:spectra_shock}, a dynamic spectrum RV\,Tau around the He\,I-5876 line. The spectra are folded on the
fundamental pulsation period (39.3 days) with the zero phase corresponding to phase zero in pulsation (see top panel of Fig. \ref{figure:shock_puls_phase}). 
The He\,I-5876 line appears in emission mostly in between phases 0.5 and 0.8, which correspond to the same phases when line doubling occurs in Fig.~\ref{figure:shock_puls_phase}.
We note that the He I-5876 line does not appear in all pulsation cycles.

\subsection{Stellar parameters} \label{section:stellarparams}
To derive the stellar parameters, we first identified the HERMES spectra of DF Cyg and RV Tau that displayed minimal shock signatures in their CCFs: only those spectra with non-deformed single-peaked CCFs were selected. 
These spectra were Doppler corrected and were then stacked to obtain a 
high S/N spectrum, with an average S/N of 100 in the blue ($\sim$ 4000 $\AA$) and 250 in the red ($\sim$ 6500 $\AA$) for both stars.

These template spectra were then normalised to the continuum level in the full HERMES spectral range ($\sim$ 3800 - 9000 $\AA$). 
We however restricted the spectral fitting to a smaller wavelength range of 5500 $\AA$ to 6200 $\AA$ because the normalisation procedure became challenging outside this range.

We used the Grid Search in Stellar Parameters (\textsc{GSSP}) software \citep{tkachenko2015} to determine the atmospheric parameters for each star. \textsc{GSSP} works by fitting 
a grid of synthetic spectra with varying $T_{\rm eff}$, $\log\,g$, $\xi$, $v_{\rm broad}$ and $[M/H]$ to the observed spectrum and outputting the $\chi^{2}$ values
of the fit. It should be noted that we quote the values for the total velocity broadening ($v_{\rm broad}$) instead of the more typical pair of the projected rotational velocity ($v \ {\rm sin} \ i$) and the macroturbulent velocity ($v_{\rm macro}$) as these parameters are often highly degenerate when the rotational velocities are low (as we expect for evolved stars). 
The synthetic spectra were generated using the \textsc{SynthV} radiative transfer code \citep{Tsymbal1996} combined with a grid of atmospheric models from the \textsc{LLmodels} 
code \citep{Shulyak2004}.

\begin{table}[h!]
  \centering
  \tiny
  \begin{tabular}{@{} llllll @{}}
    \hline \hline
    Star & Fe/H & $T_{\rm eff}$ & log\,$g$ & $\xi$ & $v_{broad}$ \\
    & (dex)& (K) &  & (km s$^{-1}$)& (km s$^{-1}$) \\
    \hline
    RV\,Tau &-0.3$\pm$0.2 & 4810$\pm$200 & 0.6 (fixed) & 3.7$\pm$0.6 & 46$\pm$3\\
    DF\,Cyg & -0.1$\pm$0.2 & 4620$\pm$130 & 0.7 (fixed) & 3.8$\pm$0.6 & 31$\pm$3 \\
    \hline
  \end{tabular}
  \caption{Stellar parameters of RV\,Tau and DF\,Cyg from spectroscopy.}
  \label{table:stellar_params}
\end{table} 

We noticed after a number of initial fitting attempts that $\log\,g$ was highly degenerate in the fit.
We therefore decided to fix $\log\,g$ to a value obtained using the following procedure: the luminosities derived from the period-luminosity-colour (PLC) relation (see Section \ref{section:luminositydf}) and 
the effective temperature obtained from the spectral fit were used to estimate the radii of the stars. 
This radius is then used to compute the surface gravity assuming a core mass for their corresponding stellar luminosity, based on the stellar models of \citet{millerbertolami2016}.

\begin{figure*}
   \centering
   \includegraphics[width=12cm,height=12cm]{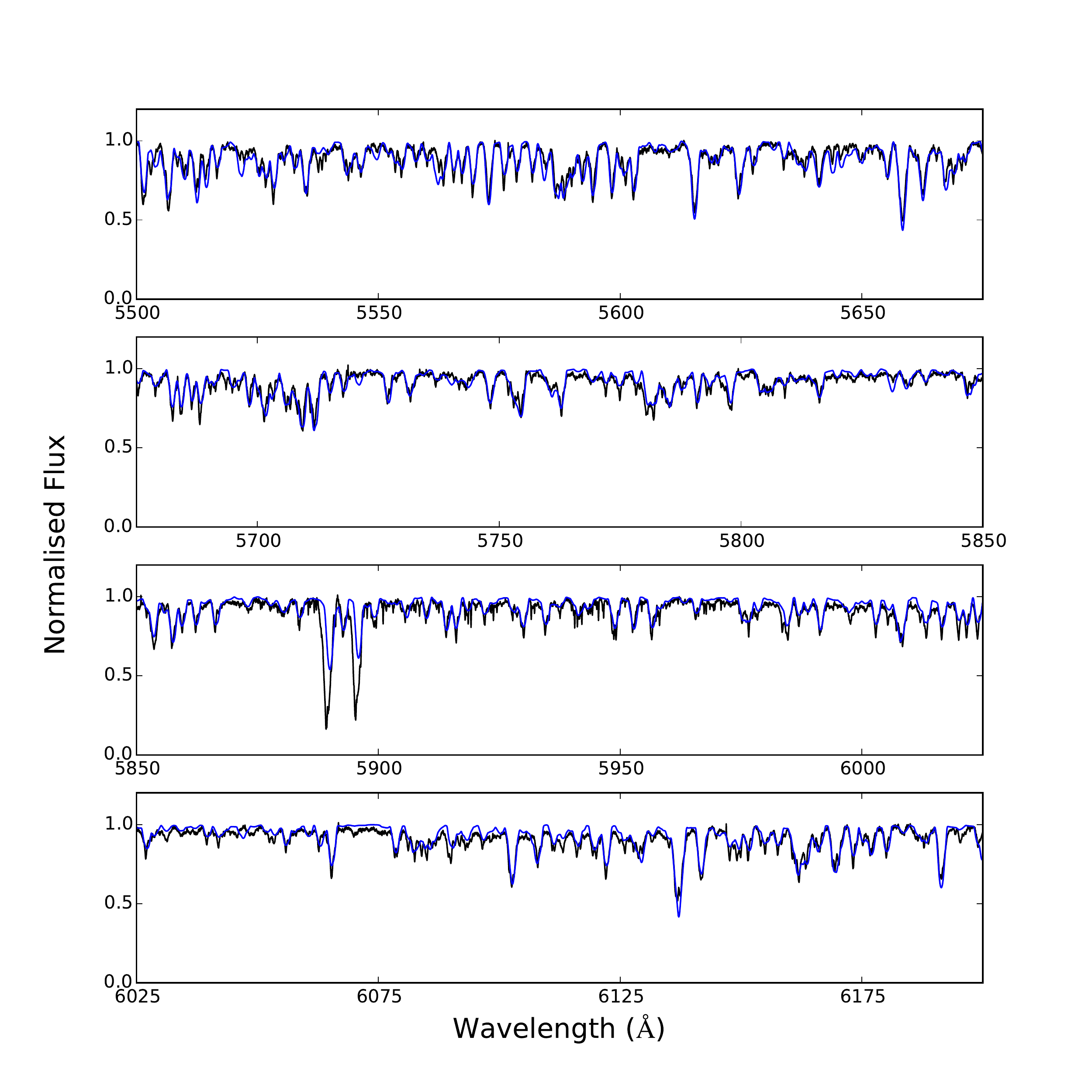}
   \includegraphics[width=12cm,height=12cm]{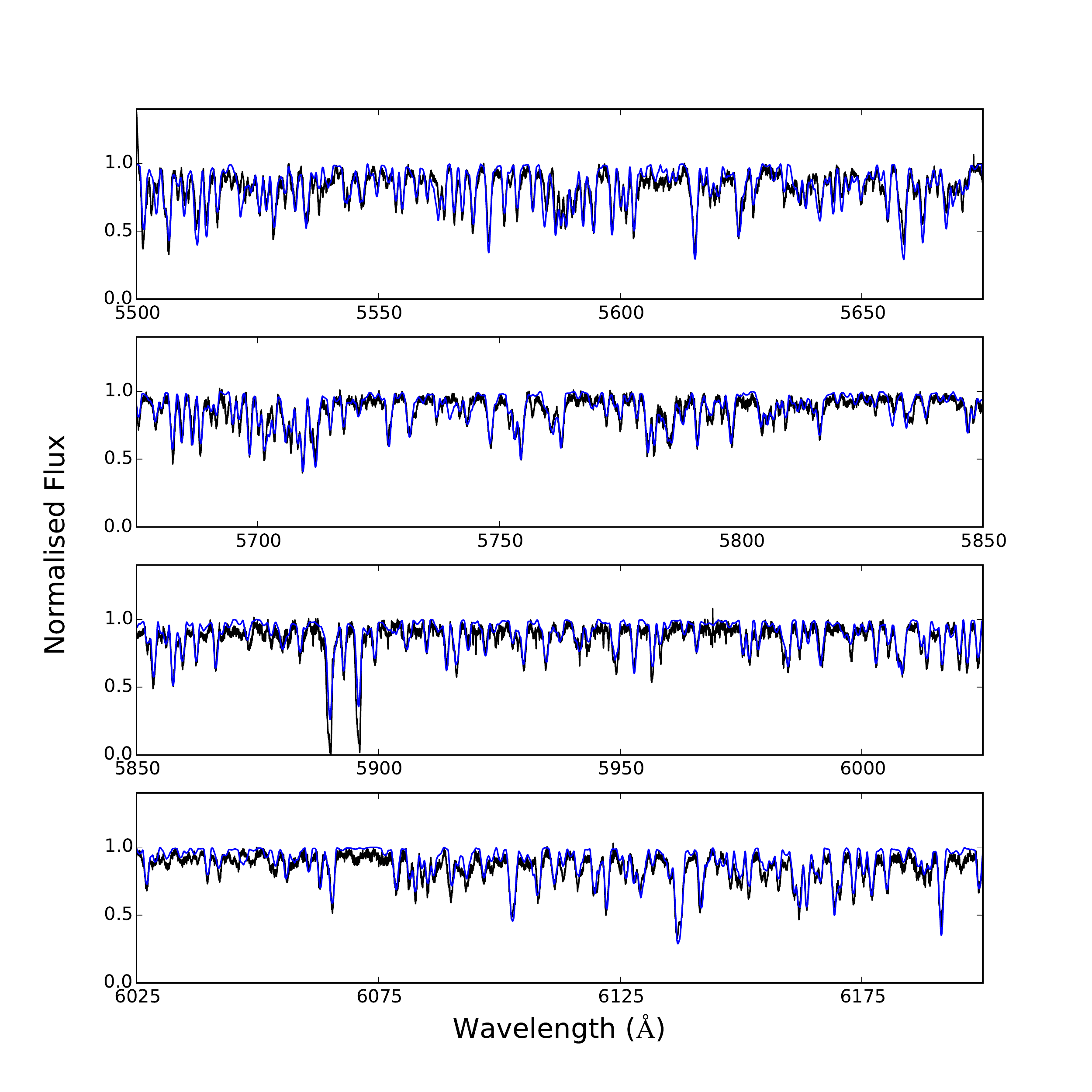}
   \caption{Spectra fitted using a Kurucz LTE model (upper panel: RV\,Tau and lower panel: DF\,Cyg). The observed spectrum is plotted in 
   black and the stellar model is plotted in blue. We have displayed chunks of 175 $\AA$ in this plot for the sake of clarity.}
     \label{figure:spectrum}
\end{figure*}

The best-fit parameters and the corresponding 1-$\sigma$ error bars are listed in Table \ref{table:stellar_params} for each star. A comparison of the best-fit 
synthetic spectra with the observed spectra is displayed in Fig. \ref{figure:spectrum}.

\section{Luminosities} \label{section:luminositydf}
The photospheric luminosities for our stars were estimated using two methods: 1) the Period-Luminosity-Colour (PLC) relation calibrated specifically 
for RV Tauri stars \citep{manick2018} and 2) the observed SED using the distance as given by the Gaia data release 2 (DR2) parallax \citep{luri2018}. 

\subsection{PLC}
The PLC calibration was based on the OGLE-III data for RV Tauri stars in the LMC \citep{soszynski08} because of its known distance. 
It was shown by \citet{manick2017} that this relation can be extrapolated to find the 
Galactic stars' luminosities because there is no metallicity dependency of the PLC relation \citep{nemec94}. The calibrated PLC relation has the form:

\begin{equation} \label{equation:eqplc}
M_{\rm bol,{PLC}} = -3.75\,\log(P_0) + 19.04 -\mu + BC + 2.55\,{(V - I)_0}\,,
\end{equation}

\noindent where $M_{\rm bol,{PLC}}$ is the bolometric absolute magnitude, $P_0$ is the fundamental pulsation period and $\mu$ is the distance modulus of the LMC, for which we adopted a value of 18.48 \citep{storm2004}. 
The bolometric correction (BC) is a function of the stellar temperature, which was computed using the calibration of \citet{flower96} and the $T_{\rm eff}$ we derived from spectroscopy. 

We calculated $E(V-I)$ using the conversion relation: $E(V-I) = 1.38\,E(B-V)$, given by \citet{tammann03} and \citet{haschke2011}.
$E(B-V)$ is the total reddening towards the star, which we obtained from the SED fitting. The observed $(V-I)$ colour of RV\,Tau was obtained from the TASS photometric catalogue \citep{droege2007} 
and from the ASAS Catalog of Variable Stars in the {\it Kepler} Field for DF\,Cyg \citep{pigulski2009}. $E(V-I)$ and $(V-I)$ for each star were used to obtain their respective intrinsic colour, ${(V - I)_0}$ which was used in 
Eq. \ref{equation:eqplc} to obtain the absolute bolometric magnitude and the luminosity of each star from the PLC ($L_{\rm PLC}$). The results are shown in Table \ref{table:lums}.

\subsection{SED} \label{section:sedfitting1}

To obtain the luminosities from the SED, we first fitted the photospheric part of the SED of both stars. The fitting was based on a parameter-grid search to reach an optimised Kurucz
model \citep[see][for details]{degroote2011}. The stellar parameters which are optimised in the fitting are: effective temperature ($T_{\rm eff}$), metallicity
[Fe/H], reddening $E(B-V)$, and surface gravity ($\log g$). 

\begin{figure}
\centering
 \includegraphics[width=9cm,height=7cm]{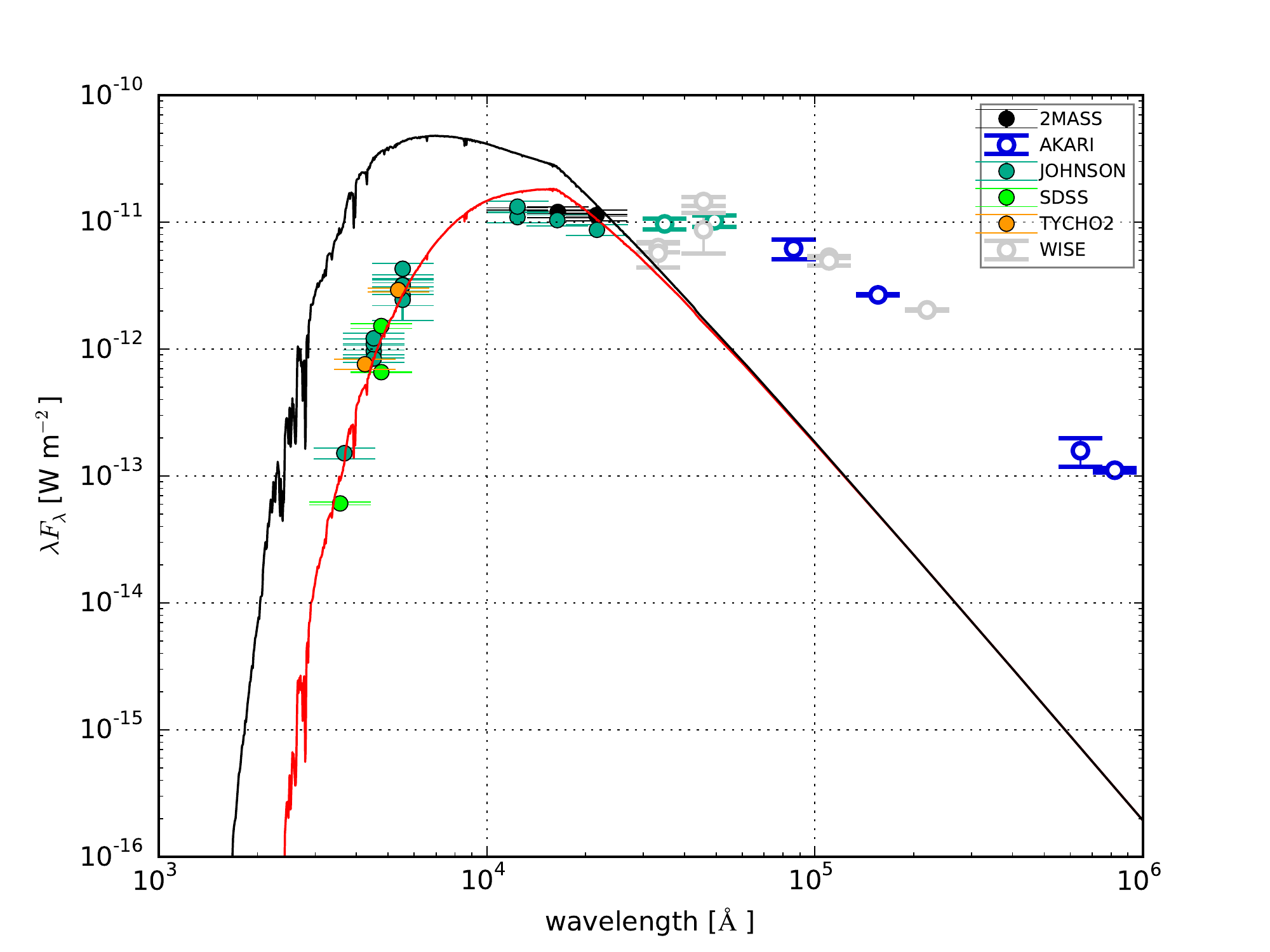}
 \includegraphics[width=9cm,height=7cm]{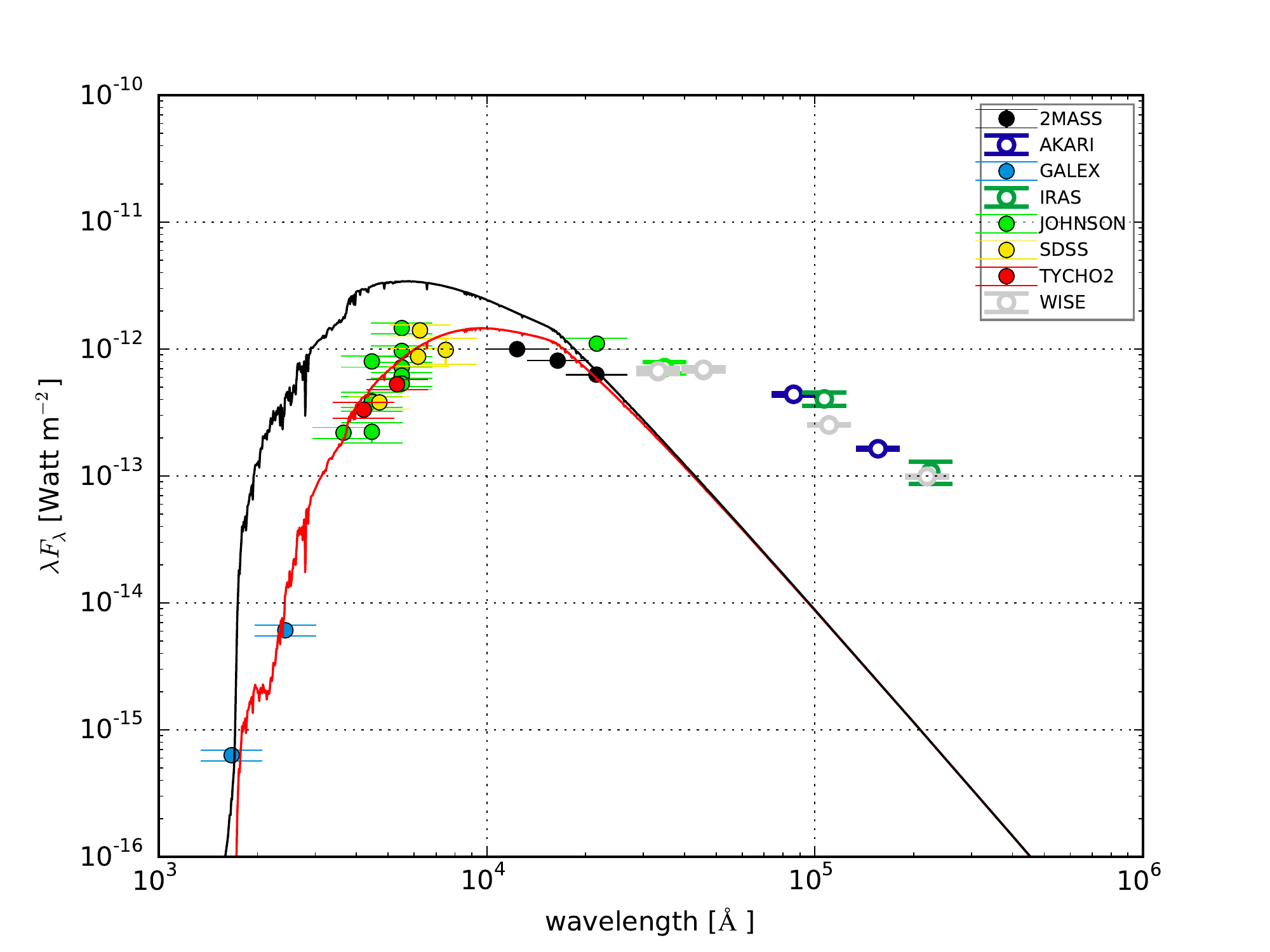}
\caption{The fitted SEDs of RV\,Tau (top) and DF\,Cyg (bottom). The red and black lines are the reddened and de-reddened models, respectively.}
\label{figure:seds}
\end{figure}

RV\,Tau and DF\,Cyg are high-amplitude pulsators and the pulsations induce a significant variance in the photospheric part of the SED, 
which is more visible in the SDSS and Johnson fluxes (see Fig. \ref{figure:seds}). Their RVb nature causes additional extinction, 
which introduces further uncertainties in the observed flux (for example, the observed 2MASS points of DF\,Cyg; see Section \ref{section:sed_data_dfcyg}).
We used the $T_{\rm eff}$, [Fe/H], $\log g$, and their respective errors that we derived from spectroscopy 
(see Section \ref{section:stellarparams}) as starting parameters in the SED fitting, with the individual errors as the bounds. 
In the SED fitting, $E(B-V)$ was left as a free parameter and
a dereddened SED model was computed for both stars using an extinction value corresponding to the lowest $\chi^2$ value. 
The fitted SED models are shown in Fig. \ref{figure:seds}.

We calculated the integrated flux ($F_{\rm int}$) using the photospheric model, fitted to the dereddened optical fluxes.
We computed the stellar luminosities from the SED using the distance from the Gaia parallax. 
We note that the Gaia DR2 parallaxes have been computed using single-star solutions and might thus not be correct in the case of binaries.
\citet{pourbaix00} have shown, however, that an unrecognised orbital motion alters the parallax derived from a single-star model only when the orbital period
is close to 1 year, which is not the case for the two stars under consideration.

The derived SED and PLC luminosities are displayed in Table \ref{table:lums}. 
The luminosities derived from both methods are in good agreement within their respective errors.

\begin{table}[ht]

\vspace{1.5ex}
\begin{tabular}{l|ll} \hline\hline\rule[0mm]{0mm}{4mm}
                              & RV\,Tau    & DF\,Cyg  \\
  \hline                                                     
$BC_V$ (mag.)                     &  -0.35     & -0.17        \\
 $E(B-V$) (mag.)              &   0.8$\pm$0.2     &  0.4$\pm$0.1     \\
($V-I$)$_0$ (mag.)             &   0.7$\pm$0.2      &   0.8$\pm$0.1    \\
Gaia dist. (kpc)              &   1.39$^{+0.13}_{-0.10}$     &  2.53$^{+0.20}_{-0.17}$   \\
$L_{\rm PLC}$ (L$_\odot$)     &  3380 $\pm$ 570   &  990 $\pm$ 190    \\
 $L_{\rm SED}$ (L$_\odot$)    &  2800$^{+430}_{-370}$       &  1010$^{+150}_{-140}$    \\
\hline
\end{tabular}
\caption{The derived luminosities from the PLC and the SED using Gaia distances using the prescription \citet{bailerjones2018}.}
 \label{table:lums}

\end{table}  

\section{Orbital analysis} \label{section:orbitalanalysis}
The high pulsational amplitude of both stars introduces a large scatter in the radial velocities, making orbital detection difficult (see e.g., Fig. \ref{figure:shock_puls_phase}).
In our previous study, we showed that these pulsations can significantly affect the parameter values derived from the Keplerian model fit \citep{manick2017}.
Thus, in RV\,Tauri stars, it is crucial to clean the radial velocities from pulsations in order to obtain accurate orbital parameters.
This method is described in the next section.

\subsection{Disentangling pulsations}
We first fit an initial Keplerian model to the original radial velocities of both stars using the orbital period obtained from the Lomb-Scargle periodogram. The fit residuals are then analysed 
to find additional periodicities which are indicative of pulsations. As can be seen in Fig. \ref{figure:shock_puls_phase}, the pulsations in the radial velocities are not purely sinusoidal. Therefore,
for both stars, we fit the best model which characterizes the pulsational behaviour. This is done by fitting a sinusoid model that includes harmonics found in 
both photometric and spectroscopic time series. 

For RV\,Tau, the pulsation periods found in the radial velocities were checked consistently against those having S/N > 4 in the ASAS photometry.
The photometric periods that matched the radial-velocity ones are marked with an asterisk ($^*$) in Table \ref{table:RV_Tau_ASAS}. 
These periods were fit using a sinusoidal model including the harmonics listed in Table \ref{table:RV_Tau_ASAS}
and the radial velocities of the sinusoid fit were subtracted from the original radial velocities to obtain a clean Keplerian orbit.

\begin{figure*}
   \centering
   \includegraphics[width=9cm,height=7cm]{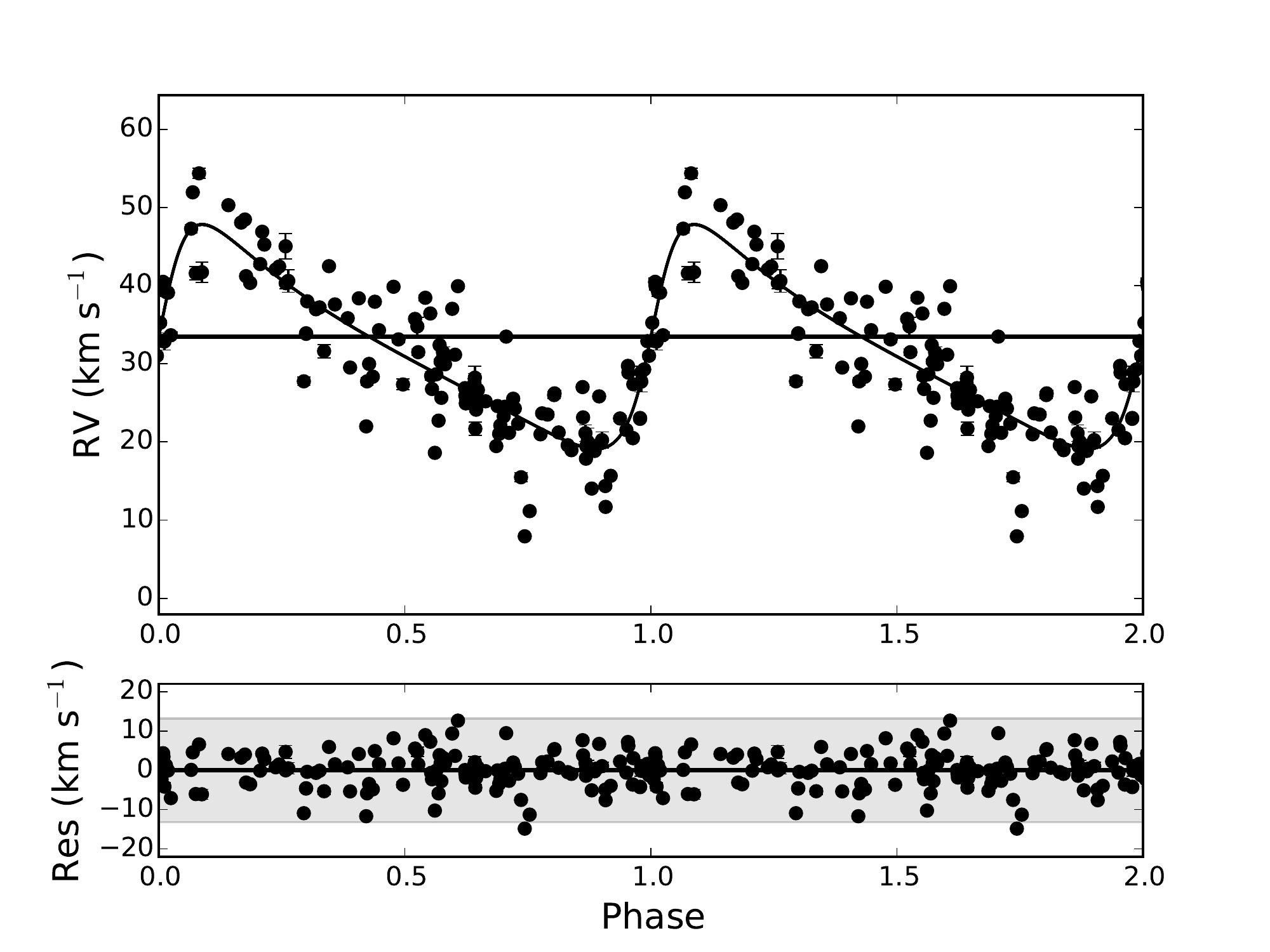} 
   \includegraphics[width=9cm,height=7cm]{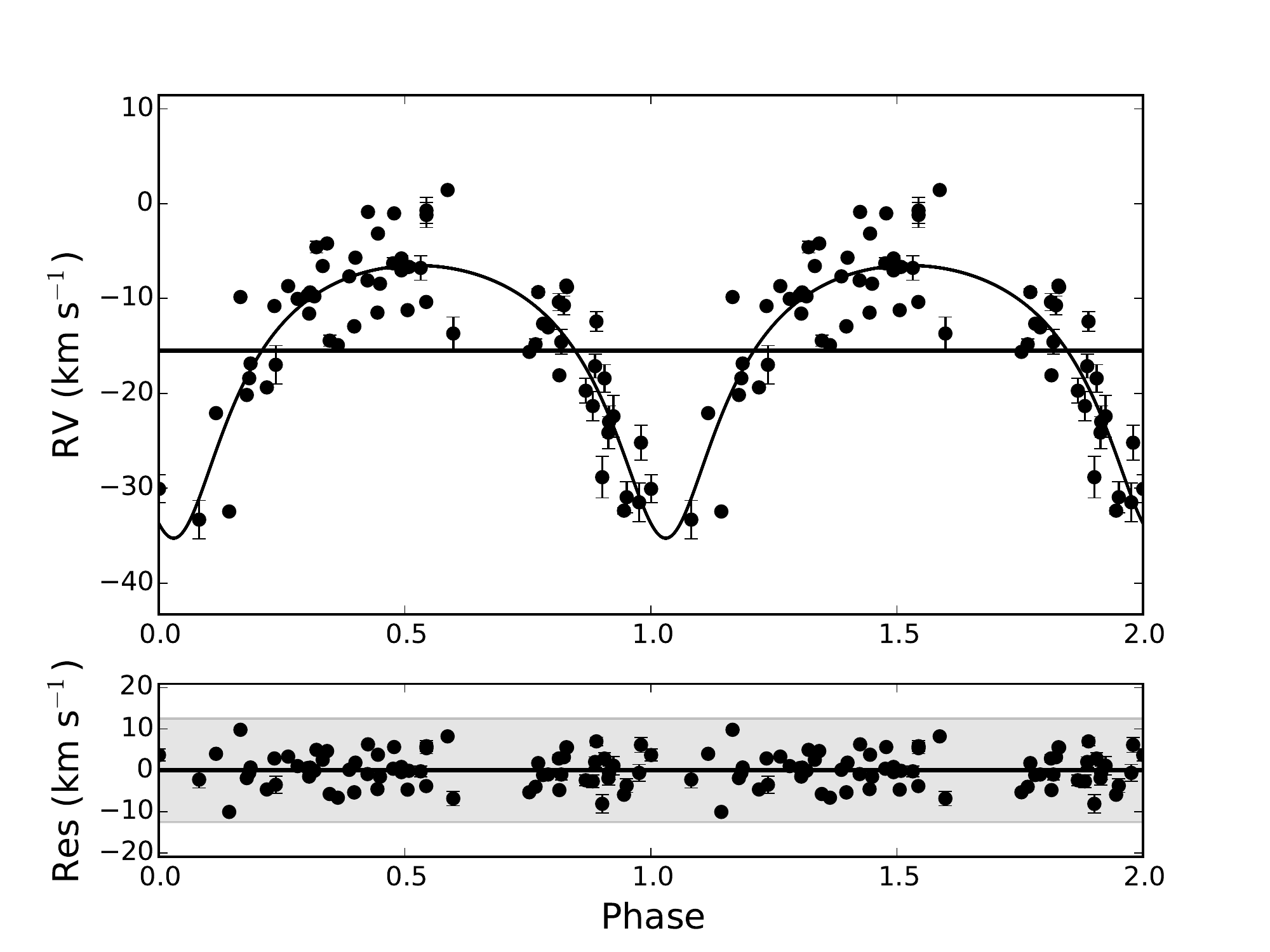} 
   \caption{Left: Cleaned orbit of RV\,Tau. Right: Cleaned orbit of DF\,Cyg. The plots are phased such as periastron passage is at phase 0.
   The orbital fit is represented by the black line and the grey-shaded regions 
   in the residuals extend $\pm$\,3$\sigma$ from the mean. }
     \label{figure:orig_clean_dfcygorbit}
\end{figure*}

The same procedure was applied to DF\,Cyg, but due to a smaller number of radial-velocity data, the only period that appeared in both photometry and spectroscopy was the fundamental period (24.9\,d).
This period was fit using a sinusoidal model with one harmonic (2$\times$24.9\,d) and one subharmonic (0.5$\times$24.9\,d). The model was then subtracted from the original radial velocities to obtain a clean orbit. 
The cleaned orbits are shown in Fig. \ref{figure:orig_clean_dfcygorbit} and the orbital parameters derived from the cleaned orbits are listed in Table \ref{table:orbitalparameters}.

\renewcommand{\tabcolsep}{0.2cm}
\begin{table*}[ht]
     \begin{small}
\vspace{0.5ex}
\begin{center}
\begin{tabular}{l|llll} \hline\hline\rule[0mm]{0mm}{4mm}
                            & RV\,Tau   &  $\sigma$ & DF\,Cyg & $\sigma$  \\
  \hline                                                     
P (days)                    &  1198     & 17    & 784          & 16 \\
K (km s$^{-1}$)             &   14.5    &  1.6   &  14.4       & 1.5 \\
e                           &  0.5      & 0.1    &  0.4         &  0.1\\
T$_0$ (BJD)                     & 2455114 & 12    &  2456195  & 23   \\
$\gamma$ (km s$^{-1}$)      &   32.0   & 0.8    &  -15.1      &  0.7  \\
$\omega$ (rad)              &   4.9   & 0.3    &  3.2      &  0.6  \\
$a_1 \sin i$ (au)           &  1.38     & 0.08    &  0.95       & 0.1   \\
f(m) (M$_{\odot}$)          &   0.24   & 0.02    & 0.19        &  0.05 \\
m$_2$ (M$_{\odot}$)          &   0.7   & 0.1    & 0.6        &  0.1 \\
N                            &   105     &  -   &  68          &  -    \\
\hline
\end{tabular}
\caption{Derived orbital parameters for the pulsation-cleaned orbits. The errors were computed using a Monte Carlo method on the data points (see text). 
N denotes the number of spectra used in the orbital analysis.}
 \label{table:orbitalparameters}
\end{center}  
\end{small}
\end{table*}  

Errors on the orbital parameters were computed using a Monte Carlo method. The data points were randomly distributed along a Gaussian with 
a mean value corresponding to the measured radial-velocity and a standard deviation equal to the standard deviation in the residuals.
The uncertainty on the orbital paramaters were obtained from the standard deviation on the set of orbital paramaters generated after 1000 Monte Carlo simulations.

\subsection{Estimating the orbital inclination}
\begin{figure*}
   \centering
   \includegraphics[width=12cm,height=9cm]{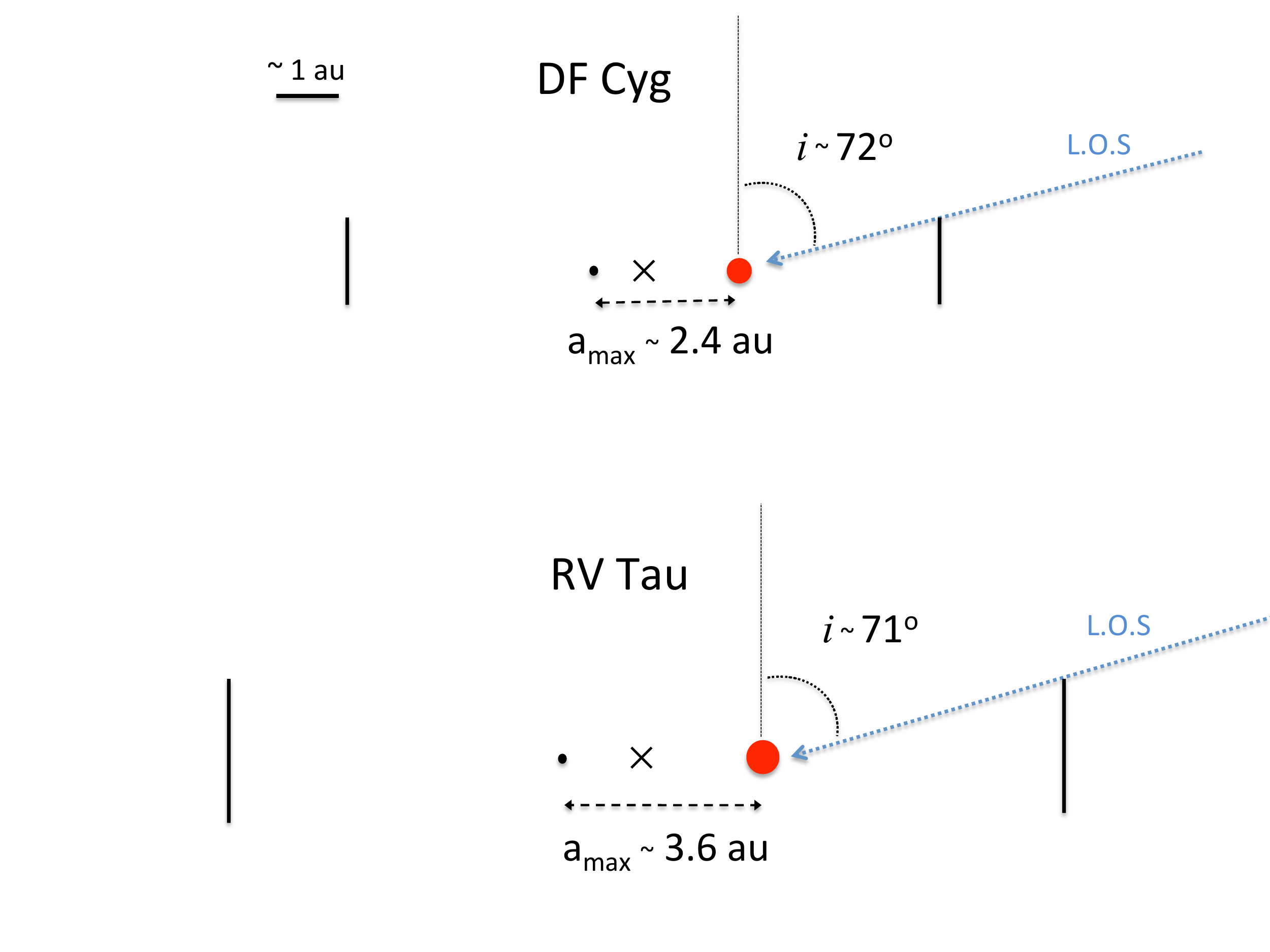}
   \caption{A scaled geometrical representation of DF\,Cyg and RV\,Tau. The dashed blue line represents our line of sight.
   The cross represents the approximate position of the centre of mass of the system based on the masses derived. 
   The size of the companion shown by the black dot, is not to scale since its radius is unknown. The vertical lines represent the inner-rim of the disc obscuring the light from the primary star.}
     \label{figure:df_cyg_geom}
\end{figure*}

The RVb phenomenon is detected in a configuration such that our line-of-sight grazes the disc (Fig. \ref{figure:df_cyg_geom}).  
The geometrical parameters of the disc inner rim and the fraction of the stellar luminosity reprocessed into IR flux by the disc can be used to constrain
the orbital inclination of the system. We assume in our analysis that the orbital plane of the binary and the disc are coplanar.

The amount of flux radiated by the dust ($F_{\rm IR}$) is a fraction of the total flux radiated by the star ($F_{\rm star}$).
The ratio $F_{\rm IR}$/$F_{\rm star}$ thus represents the fraction of the total flux that is being absorbed and re-emitted by the dust. It can be estimated from the SED. 

We computed $F_{\rm star}$ from the dereddened SED model. The amount of flux emitted by the dust was obtained by integrating the IR part of the SED and subtracting the stellar contribution.
In this process, we assumed that all the IR-excess is of thermal origin and that there is no scattering of the optical photons affecting the optical fluxes. 

We assume that the flux radiated from the star is impacting on the inner rim of the disc. Its height ($h$) is defined as the height above which stellar photons
can pass without being absorbed. It can be shown 
that the ratio $F_{\rm IR}$/$F_{\rm star}$ scales as $h/2R$, where $R$ is the inner-rim radius of the disc.
The inner boundary of the dust disc is set by the dust sublimation temperature ($T_{\rm sub}$). Assuming that 
the inner rim is mostly composed of silicates, we adopt $T_{\rm sub}$ to be $\sim$ 1300 K \citep[e.g.,][]{kama2009,hillen2016}.
The dust sublimation temperature $T_{\rm sub}$ is a function of the inner-rim radius ($R$), luminosity ($L_{\rm star}$) through the relation \citep{dullemond01,kama2009}: 

\begin{equation} \label{equation:innerrim}
R = \frac{1}{2}\,\left(\frac{C_{\rm bw}}{\epsilon}\right)^{1/2}\,\left(\frac{L_{\rm star}}{4\pi\,\sigma\,T_{\rm sub}^4}\right)^{1/2}\,,
\end{equation}

\noindent where $C_{\rm bw}$ is the backwarming coefficient, ranging from $\sim$ 1 to 4 depending on the optical thickness of the disc, the lower bound corresponding to
an optically thin disc \citep{kama2009}. 
The cooling efficiency ($\epsilon$) of the dust grains is defined as the ratio $\epsilon$ = $\kappa(T_{\rm dust})$/$\kappa(T_{\rm star})$, where $\kappa$ is the Planck mean opacity of the dust
species at the specified temperature \citep{kama2009}. 
$C_{\rm bw}$ and $\epsilon$ are highly dependent on the intrinsic properties of the dust around the stars. Given that we have no \textit{a priori} knowledge of the dust 
composition around these two stars, we assumed $\epsilon \sim$1 for both of them \citep[e.g.,][]{lazareff2017}. In the case of DF\,Cyg, the large change in optical thickness through the 
RVb cycle suggests that the disc is optically very thick in the radial direction (see Section \ref{section:extinctionvariation}). 
We therefore assigned a value of $C_{\rm bw}$ = 4 for DF\,Cyg.

$L_{\rm star}$ in Eq. \ref{equation:innerrim} is the stellar luminosity ($L_{\rm SED}$), listed for both stars in Table \ref{table:lums}.
The inner-rim radius and the fraction $F_{\rm IR}$/$F_{\rm star}$ were used
to estimate the height $h$ of the disc inner rim. The inner-rim radius and $h$ were used to compute the inclination using simple geometry
and assuming that the disc is fully optically thick. The results are displayed in Table \ref{table:inclinations}.

\begin{table}
  \centering
  \tiny
  \begin{tabular}{@{} lllllll @{}}
    \hline
    Star & $L_{\rm IR}$/$L_{\rm total}$ & $T_{\rm sub}$ & $R$    & $h$  & Inclination ($i$)\\
         &                      & (K)        & (au)   & (au) & ($^\circ$) \\
    \hline \hline
    DF\,Cyg & 0.27             & 1300 & 3.1$\pm$0.6 & 1.0$\pm$0.1 & 72$\pm$6 \\
    RV\,Tau & 0.39             & 1300 & 4.9$\pm$0.8 & 1.6$\pm$0.3 & 71$\pm$8 \\
    \hline
  \end{tabular}
  \caption{Disc parameters and orbital inclinations.}
  \label{table:inclinations}
\end{table}

\subsection{Asymmetry in the RVb phenomenon}
\begin{figure*}
   \centering
   \includegraphics[width=12cm,height=10cm]{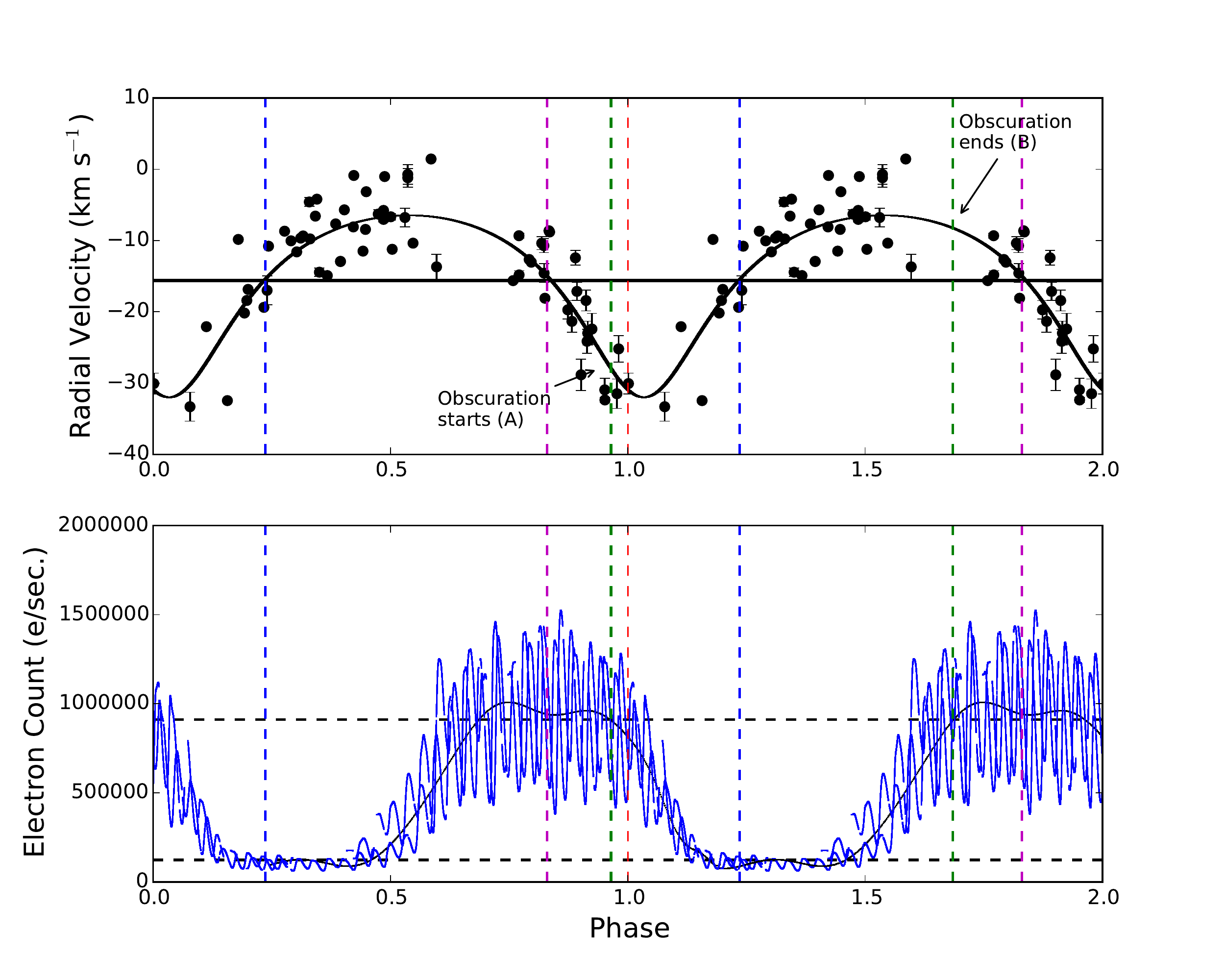}
   \caption{Top: Spectroscopic orbit of DF\,Cyg. The vertical blue and magenta dashed lines show inferior and superior conjunctions, respectively.
   The red dashed vertical line shows the phase of periastron passage and the vertical green dashed lines show the phases at which the obscuration start and end.
   Bottom: The occurrence of the same events in the \textit{Kepler} light curve. The black line is an interpolation through the light curve. The horizontal dashed lines are the mean flux at maximum and minimum light phases of the RVb cycle.}
     \label{figure:dfcyg_rvs_flux_time}
\end{figure*}

We notice an asymmetry in the RVb phenomenon; the long-period minimum is not centered symmetrically around the occurrence of the inferior conjunction (see Fig. \ref{figure:dfcyg_rvs_flux_time}).

To estimate the occurrence of the start and end of the obscuration, we first computed the mean flux at maximum and minimum light phases of the RVb cycle.
These mean fluxes are represented by the upper and lower dashed horizontal lines in the lower panel of Fig. \ref{figure:dfcyg_rvs_flux_time}. 
The start and end of the obscuration (points A and B, respectively) are shown by the vertical green dashed lines and have been defined as the points at which these mean fluxes
intersect the fitted polynomial to the lightcurve (see black line in bottom panel of Fig. \ref{figure:dfcyg_rvs_flux_time}). 

\begin{figure}
   \centering
   \includegraphics[width=9cm,height=7cm]{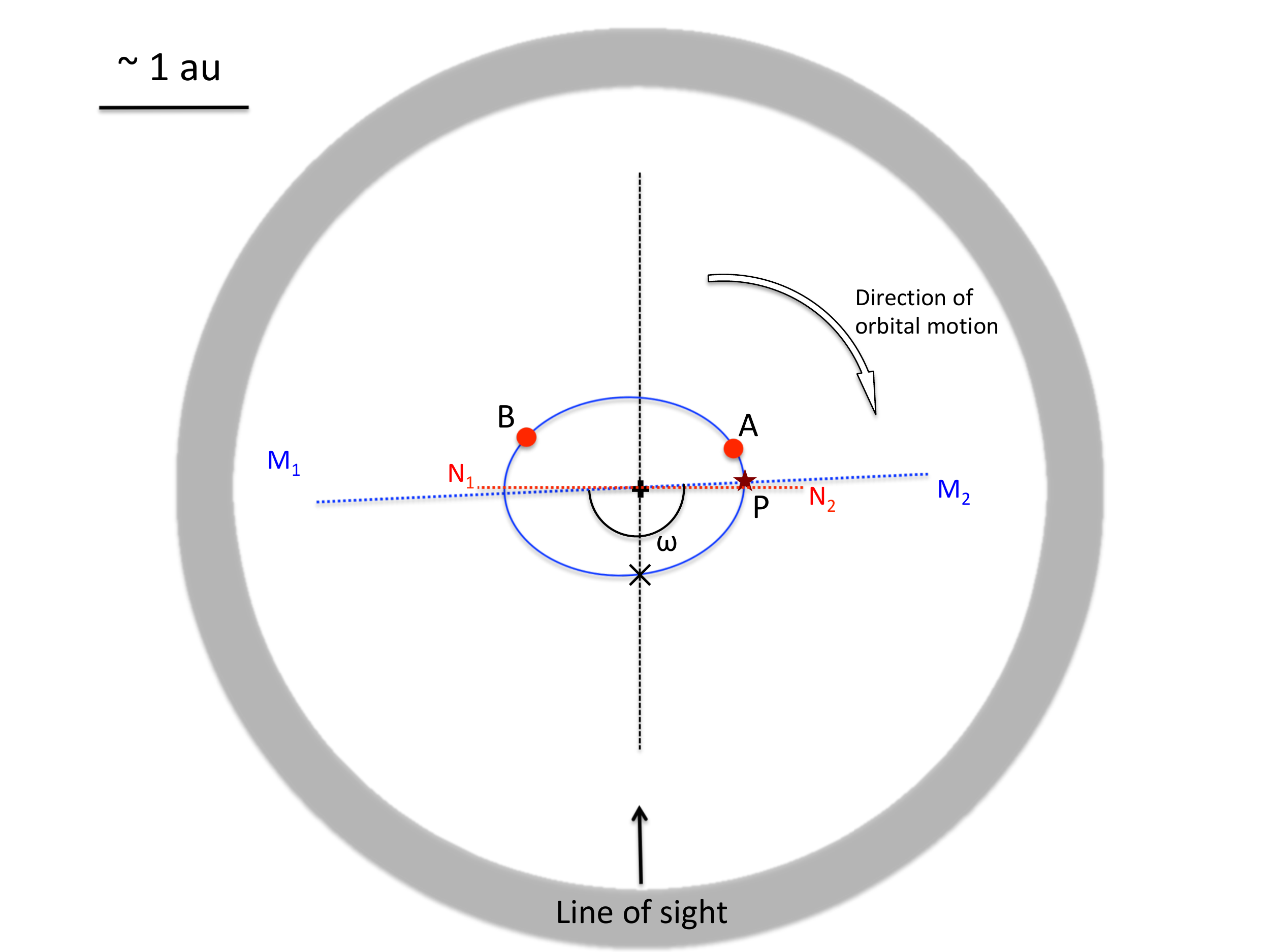}
   \caption{An approximate scaled top view representation of the binary system DF\,Cyg.
   The orbit of the primary star is shown as the blue ellipse. The position of the primary star at start and end of obscuration is shown 
   as red filled circles at positions A and B, respectively. The periastron passage is marked with a $\star$ labelled P, 
   the inferior conjunction as a $\times$ symbol and the center of mass of the binary system (and of the disc) is marked with a $+$ symbol. 
   The line of nodes is represented by the line N$_1$, N$_2$ and the major axis of the orbit is represented by the line M$_1$, M$_2$. 
   The longitude of periastron, $\omega$, has a value of 183$^\circ$ as obtained from the orbital fit.
   The grey shaded region represents the circumbinary disc.}
     \label{figure:df_cyg_ecc}
\end{figure}

We display in Fig. \ref{figure:df_cyg_ecc} a scaled geometrical representation of the system to visualise this asymmetry better. 
The line [M$_1$, M$_2$] shows the major axis of the orbit and line [N$_1$, N$_2$] is the line of nodes, N$_1$ and N$_2$ being the ascending and descending nodes, respectively. 
The longitude of periastron, $\omega$, has a value of 183$^\circ$.  
The periastron passage is marked as P and the inferior conjunction is marked as a cross. Points A and B in Fig. \ref{figure:df_cyg_ecc} show the positions
at which the obscuration of the primary star starts and ends, respectively.

The asymmetry with respect to inferior conjunction can be explained by the fact that the 
primary star is obscured for a longer period of time on one side of the inferior conjunction 
than on the other side. This is better visualised in the geometric representation (Fig. \ref{figure:df_cyg_ecc}), whereby the star 
spends a longer time from inferior conjunction to point B than point A to inferior conjunction.

\begin{figure} \label{rvtau_inf_sup}
   \centering
   \includegraphics[width=9cm,height=7cm]{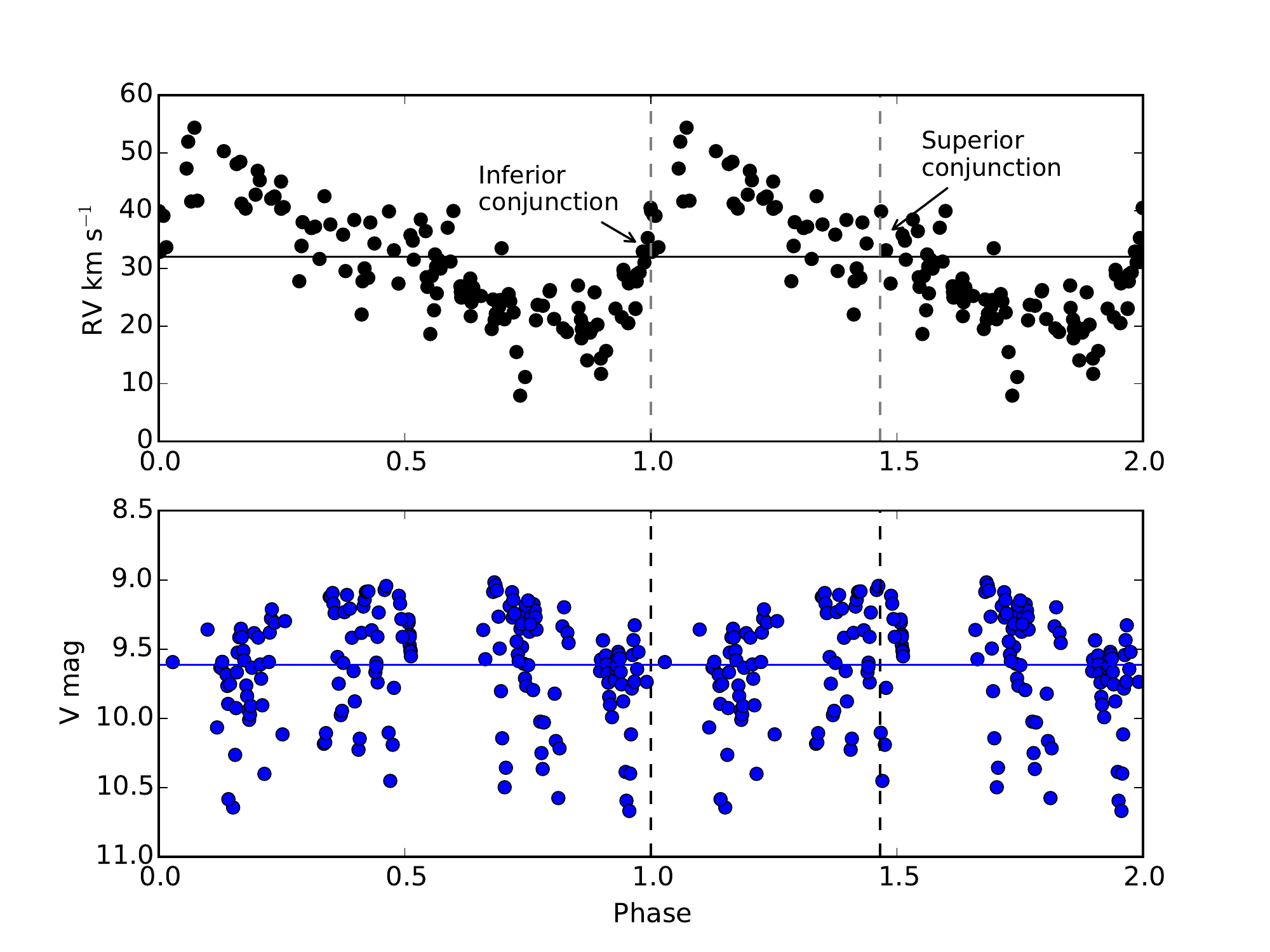}
   \caption{Top: The spectroscopic time series of RV\,Tau phased on the orbital period.
   Bottom: The photometric time series of RV Tau phased on the orbital period. The horizontal lines show the mean velocity (top) and $V_{\rm mag}$ (bottom).
   The vertical dashed lines indicate the phases at which inferior and superior conjunction occur.
   }
     \label{figure:rvtau_inf_sup}
\end{figure}

We carried out the same procedure for RV\,Tau to show that the dimming and brightening of the star due to dust obscuration events (see Figure \ref{figure:rvtau_inf_sup}).
The dimming of the star already starts at a phase prior to 
inferior conjunction denoted by the vertical dashed line at phase $\sim$ 1.0
and it starts brightening before reaching superior conjunction. 

\subsection{Masses}
Knowledge of the orbital inclinations of the systems via the RVb phenomenon enables us to derive accurate masses of the companions.
For this purpose, we use the mass function obtained from the fitted Keplerian orbit to calculate the companion mass ($m_2$) assuming a primary 
mass ($m_1$). The mass function is given by:

\begin{equation} \label{equation:massfunct}
 f(m_1,m_2)=\frac{m^3_{2}}{(m_{1}+m_{2})^2} \sin^3 \textit{ i}=\frac{P}{2 \pi \textit{G}} K^3_{1}
\end{equation}

\noindent where $K_1$ is the semi-amplitude of the primary. Given that we know the luminosities (\ref{table:lums}), we can use stellar evolution models to estimate the mass of the primary star since there is a strong correlation between core mass and luminosity for post-AGB stars. 
For RV\,Tau, the luminosity of 2800~L$_\odot$ gives a mass of $\sim$ 0.53~M$_\odot$ according to the models of \citet{millerbertolami2016}. Given the low luminosity of DF\,Cyg, 
it must be a post-RGB star (see Section \ref{section:evostage}). Since post-RGB tracks are not available in the literature, we have used the \textsc{MESA} code \citep{paxton11} to compute post-RGB models. This was done 
by artificially increasing mass loss at different moments during the RGB phase in order to remove the stellar envelope. We found that a luminosity of $\sim$ 1010~L$_\odot$ corresponds to a core mass of $\sim$ 0.4~M$_\odot$. 
We used these primary masses and the derived orbital inclinations to compute the companion masses for each star. The results are displayed in Table~\ref{table:orbitalparameters}. 

\section{Discussion} \label{section:discussiondf}

\subsection{Evolutionary stage} \label{section:evostage}
In our recent studies, we have shown that around 25\% of all the known RV Tauri stars display a broad IR excess in their SEDs \citep{gezer2015,manick2018}. 
It is well established that the nature of the circumstellar IR excesses shows the presence of dust in the form of a Keplerian disc around the stars. 
It was shown in \citet{manick2018} that RV Tauri stars with a dusty disc 
can either be post-AGB or post-RGB stars, the post-RGB stars being a new class of objects that have similar spectroscopic 
stellar parameters as post-AGB stars, but they appear at lower luminosities on the HRD \citep{kamath2015,kamath2016}.
Thus, given the strong link between the presence of a dusty disc and binarity among the Galactic RV Tauri stars, one of our 
main conclusions in \citet{manick2018} was that dusty RV Tauri stars are likely all binaries. The low luminosity ones
have likely evolved off the RGB while the higher luminosity ones are likely post-AGB objects.
Our results presented here on RV\,Tau and DF\,Cyg, further corroborates this conclusion.
With our derived luminosities and stellar parameters, we conclude that RV\,Tau is likely a genuine post-AGB 
binary, while DF\,Cyg is a post-RGB binary given its low luminosity.

\subsection{Roche-lobe filling} \label{section:rochelobefilling}
Given their non-zero eccentricities, the orbital separations in the RV Tau and DF Cyg systems change by a significant value between apastron and periastron passage. 
We have used the Eggleton formula (Eggleton 1983) to estimate the Roche-lobe size ($R_{\rm RL}$) at these two phases and hence the Roche-lobe filling factor ($f$), which is given by $f=R_{\rm AGB/RGB}/R_{\rm RL}$.
The value of $f$ allows us to investigate the efficiency of the interaction between the primary and the companion, when the primary was at giant sizes, i.e. when it is ``evolved back'' with the same luminosity to the RGB or AGB.
Assuming a typical RGB/AGB effective temperature of $\sim$ 3500 K and their derived luminosities, we estimate $f$ to be $\sim 1$ at periastron passage for both stars, 
while at apastron we obtain a value of $\sim$ 0.5 for DF\,Cyg and $\sim$ 0.6 for RV\,Tau.

The derived high Roche-lobe filling factors indicate that these stars should have indeed interacted on the RGB or AGB. 
This indicates that the disc was possibly formed during this evolutionary phase, likely via mass loss through L$_2$. 
However, this interaction did not lead to a dramatic spiral-in.

\subsection{The H$_\alpha$ profile of RV\,Tau and DF\,Cyg} \label{section:jets}
Studies of post-AGB binaries by \citet{gorlova14,gorlova15} and \citet{bollen17} showed that the H$\alpha$ profile 
appears as double-peaked at inferior conjunction with a central absorption feature, while it
turns into a P-Cygni-like profile at superior conjunction. The radial velocities related to the H$\alpha$ absorption
are seen to have projected values up to $\sim$ 300 km s$^{-1}$ in some systems. The origin of these 
high velocities is best explained by fast outflows in the form of jets launched by the companion \citep{bollen17}. In this scenario, the 
H$\alpha$ line appears as P-Cygni when the luminous primary is occulted by the jet, which is only seen 
when the primary's continuum photons are scattered out of the observer's line-of-sight. 
The P-Cygni profile only occurs when the companion is in front of the highly luminous primary.

\begin{figure}
   \centering
   \includegraphics[width=9cm,height=7cm]{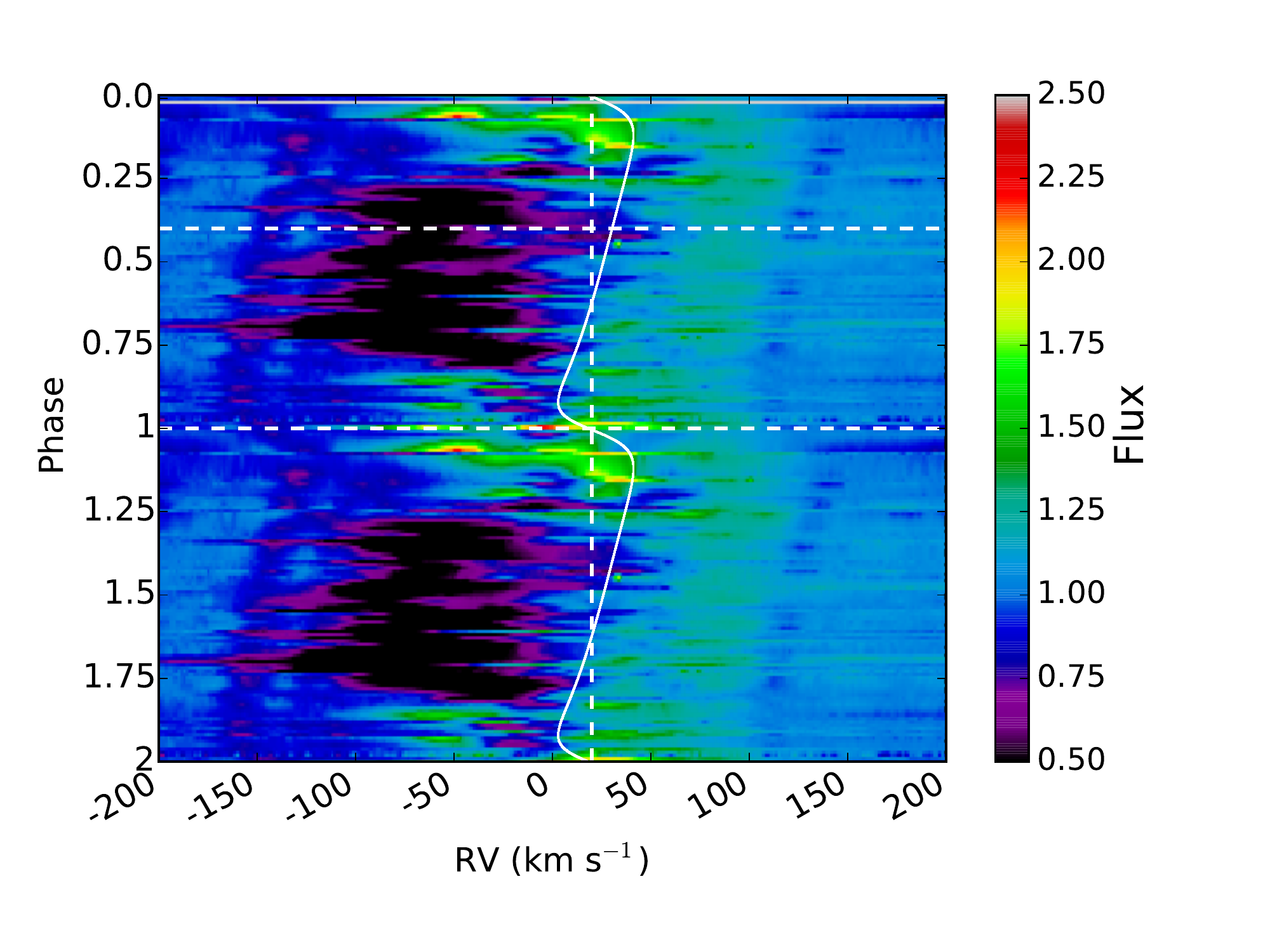}
   \includegraphics[width=9cm,height=7cm]{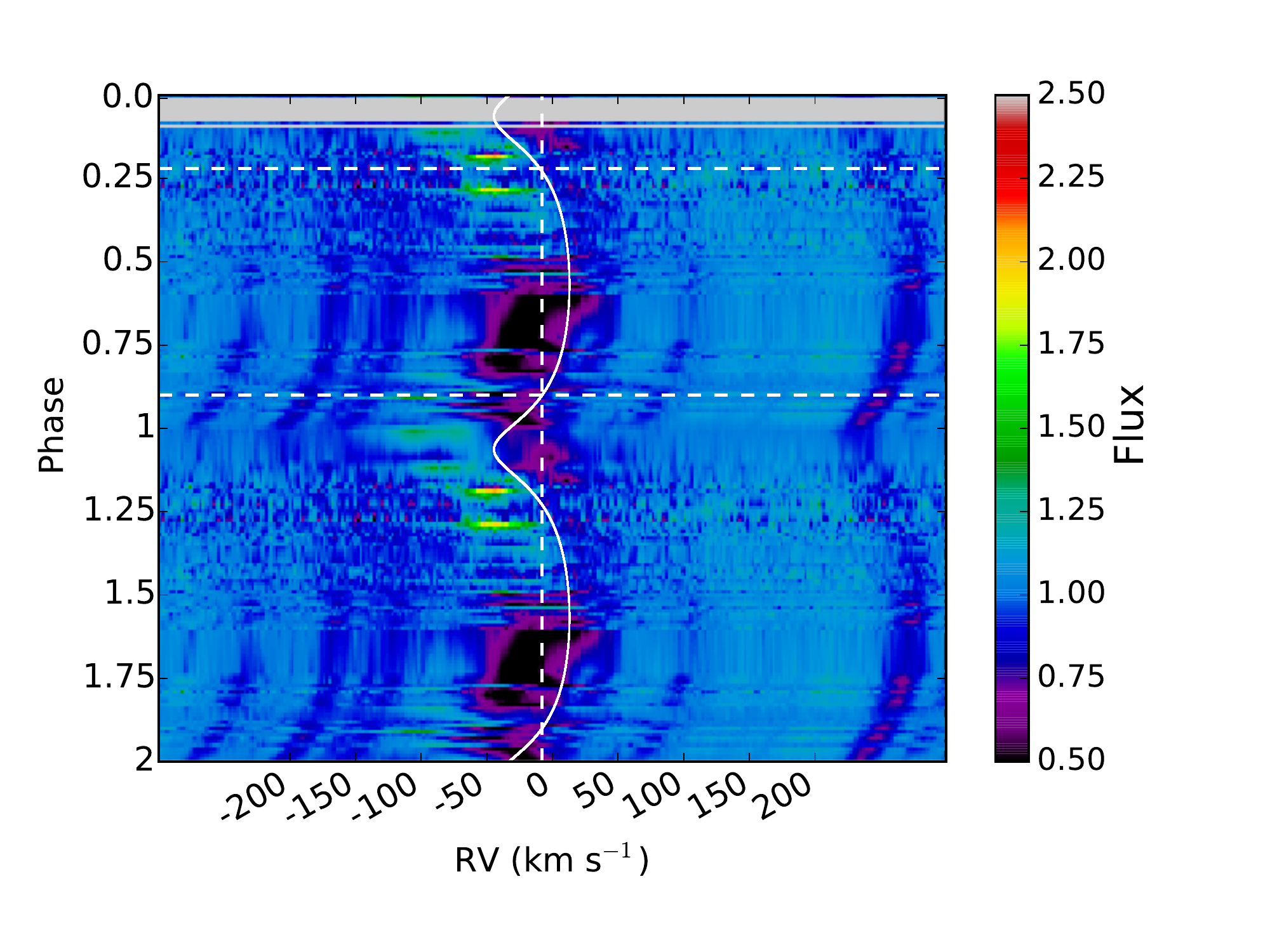}
  \caption{A dynamic spectrum of the H$\alpha$ profile variation as a function of the orbital phase for RV\,Tau (Top) and DF\,Cyg (Bottom). 
  The pulsations have been removed to better display the orbital motion. The phases in both plots correspond to phases 
  in the spectrocopic orbits shown in Figure \ref{figure:orig_clean_dfcygorbit}. The plotted curve is the orbit of the primary. The vertical dashed lines represent the systemic velocity. 
  The upper and lower horizontal dashed lines correspond to superior and inferior conjunction, respectively for RV\,Tau and vice-versa for DF\,Cyg.}
     \label{figure:rvtau_halpha}
\end{figure}

We used the pulsational radial velocities to clean the pulsations from the spectra of both stars. The spectra were then phase-folded using their respective orbital period 
to produce a pulsation-cleaned dynamic spectrum of the H$_\alpha$ profile as shown in Fig. \ref{figure:rvtau_halpha}.
In both cases, the absorption feature is seen to cover a large fraction of the orbital phase, 
suggesting that the opening angle of the outflow must be high.

In RV\,Tau, the outflow velocities are seen to be up to $\sim$ 150 km s$^{-1}$ 
with a deprojected velocity of $\sim$ 450 km s$^{-1}$, assuming a 71$^{\circ}$ inclination. The outflow is also seen to be non-symmetric 
with respect to superior conjunction (see Top panel of Figure \ref{figure:rvtau_halpha}). We suspect that the outflow from the 
companion of RV\,Tau is not orthogonal to the orbital plane. There is a tilt in the outflow which causes a lag in the absorption 
feature with respect to the superior conjunction. High angular resolution data will be needed to resolve this.

The outflow velocity in DF\,Cyg is found to be of the order of 50 km s$^{-1}$. Assuming an orbital inclination of 72$^{\circ}$, the deprojected outflow velocity is $\sim$ 160 km s$^{-1}$.
We also note that the dynamic spectrum is less smooth at inferior conjunction compared to superior conjunction (see bottom panel of Fig. \ref{figure:rvtau_halpha}).
We compared the S/N in the spectra corresponding to these two phases and find that on average the S/N is lower in spectra taken at inferior conjunction.
This is explained by the fact that since the average exposure time is the same irrespective of the orbital phase, the overall signal in spectra taken around 
inferior conjunction is low due to high circumstellar extinction during those phases. 

\subsection{Nature of the companions}
From the orbital inclinations and primary masses, we find the companion masses to be $\sim$ 0.7$\pm$0.1 M$_{\odot}$ and 0.6$\pm$0.1 M$_{\odot}$ for RV\,Tau and DF\,Cyg, respectively.
Considering there is no detectable flux contribution from the secondary in the spectra, there are two possibilities for the companion types: either a cool unevolved main-sequence star or a white-dwarf (WD). 
We argue that the companion is most likely a main-sequence 
star based on the fact that a WD would have escape velocities of the order of $\sim$ 5000 km s$^{-1}$, which is much higher than the outflow velocities seen in these objects \citep{bollen17}. A typical main-sequence star 
would have escape velocity of around 370 - 500 km s$^{-1}$ \citep{praderie86,willson87}, which is of the order of the deprojected velocities we detect in these systems.

\subsection{The RVb phenomenon}
We have shown in this work that the spectroscopic orbital period corresponds to the long-term photometric period seen in the light curves of RV\,Tau and DF\,Cyg and thus 
binarity best explains the RVb phenomenon operating in RV Tauri stars. This is complemented by the findings presented in \citet{manick2017} of three more Galactic RV Tauri stars of the RVb photometric classes being 
spectroscopic binaries with orbital periods similar to the long-term photometric variability.

The detection of the periodic extinction is, however, dependent on the orbital inclination of the system.
If the system is observed close to pole-on, there is no material in our line-of-sight to extinct the light from the star and thus no long-term variability is observed.
Conversely if the system is seen edge-on, the disc obscures the star completely and the likely RVb phenomenon will only be observed in scattered light, as in the case of the post-AGB binary HD\,44179 \citep{cohen04,kiss07}.

With the number of known disc RV Tauri stars in the Galaxy and Magellanic Clouds being $\sim$ 45, there are 15 ($\sim 33\%$) of them which display the RVb pheonomenon \citep{preston63,pollard96,gezer2015,manick2017,manick2018}.
This fraction represents the probability of RVb detection among the optically bright RV Tauri stars with a disc, assuming the plane of the discs are randomly inclined in the sky.

Using this probability, we can estimate the range of inclinations in which the variable extinction would be detected. 
Using the fact that the total probability representing all range of possible orbital inclinations (i.e. 0 - 90 degrees) must be 1,
we can compute the range of inclinations that correspond to a probability of 0.33 from the condition,

\begin{equation}
\int_{x_1}^{x_2}\sin(x)\,{\rm d}x = 0.33\,,
\end{equation}

\noindent where [$x_1$, $x_2$] represents the range of inclinations. Assuming that the optimal orbital inclination to detect the 
RVb phenomenon is $\sim$ 70 degrees, we obtain the range [$x_1$, $x_2$] to be 60 to 80 degrees.

\section{Conclusions}   \label{section:conclusiondf}  
In this paper, we have presented the spectroscopic orbits of two binary RV Tauri stars: DF\,Cygni and the eponym RV\,Tauri. 
Their derived spectroscopic orbital periods are 1198$\pm$17 days and 784$\pm$16 days, respectively, and have eccentricities of 0.5$\pm$0.1 and 0.4$\pm$0.1, respectively.
The spectroscopic orbital periods correspond to the long-term periodic variability seen in the photometric time series of these two stars. 
Our finding that the long-term photometric period corresponds to the spectroscopic orbital periods makes us
conclude that the RVb phenomenon is best explained by variable extinction by the circumbinary disc due to orbital motion. 
In this scenario, the primary star (i.e, the luminous RV Tauri star) is being extinct periodically 
by the disc due to its orbital motion and this extinction is seen in the time series as a reduction 
in the mean stellar flux and suppression of the observed pulsation amplitude. 
The phase at which the mean flux from the star is minimum corresponds to inferior conjunction.
We also find that the disc material of DF\,Cyg grazing our line-of-sight is optically very thick.

Our line-of-sight grazes the disc in both systems. Assuming that the inner-rim of the disc is at the dust sublimation temperature and that the orbit 
of the binary and the disc are coplanar, we estimate the orbital inclination to be $\sim$ 71$\pm$8 degrees for RV\,Tau and 72$\pm$6 degrees for DF\,Cyg.
Knowledge of the inclination allows accurate determination of the secondary mass if the primary mass is known.
To estimate the primary mass, we used the derived luminosities of the stars and the evolutionary models of \citet{millerbertolami2016}. 
The derived companion masses indicate that the secondary is likely a low-mass unevolved main sequence star in both cases. Additional arguments
for this is that the outflow velocity of the jet around the companion corresponds to the escape velocity we expect from a main-sequence star rather than a WD.

We have also analysed the high quality \textit{Kepler} photometry of DF\,Cyg and found evidence of modulation in the fundamental pulsation peak. The orbital period of the system can be constrained using the sidelobe features and we have shown that it corresponds to the orbital period found from both photometry and spectroscopy.

We used a combination of the pulsation periods and the Gaia distances to compute their photospheric luminosities using two methods.
The luminosities derived in both cases are in good agreement within their respective errors. Based on the luminosities, we conclude that RV\,Tau is a post-AGB binary, 
while DF\,Cyg is likely a post-RGB binary.

\begin{acknowledgements} 
The spectroscopic results presented in this paper are based on observations made with the Mercator Telescope, 
operated on the island of La Palma by the Flemish Community, at the Spanish Observatorio del Roque de los Muchachos of the Instituto de Astrof\'isica de Canarias. 
This paper also includes data collected by the {\it Kepler} mission. Funding for the {\it Kepler} mission is provided by the
NASA Science Mission Directorate. RM, DK, and HVW acknowledge support from the Research Council of K.U. Leuven under contract GOA/13/012. RM, HVW and AJ acknowledges support of the Belgian Science Policy Office under contract BR/143/A2/STARLAB. HVW acknowledges additional support from the Research Council of K.U. Leuven under contract C14/17/082, and from the Science Foundation of Flanders (FWO) under contract G075916N.
DK acknowledges support from the Macquarie University New Staff Scheme funding (MQNS 63822571).
The research leading to these results has partly been funded by the European Research Council (ERC) 
under the European Union's Horizon 2020 research and innovation programme (grant agreement N$^{o}$670519: MAMSIE).
\end{acknowledgements}

 \bibliographystyle{aa}          
 \bibliography{allreferences}

\end{document}